\documentclass[aps,prx,twocolumn,superscriptaddress,longbibliography,nofootinbib]{revtex4-1}

\usepackage{amsmath,amsfonts,amssymb}
\usepackage{braket}
\usepackage{graphicx}
\usepackage{tikz}
\usetikzlibrary{shapes,arrows,positioning,calc}
\usepackage{textcomp}
\usepackage{relsize}

\newcommand{\be}{\begin{equation}}
\newcommand{\ee}{\end{equation}}

\usepackage[hidelinks]{hyperref}

\AtBeginDocument{%
    \newwrite\bibnotes
    \def\bibnotesext{Notes.bib}
    \immediate\openout\bibnotes=\jobname\bibnotesext
    \immediate\write\bibnotes{@CONTROL{REVTEX41Control}}
    \immediate\write\bibnotes{@CONTROL{%
    apsrev41Control,author="08",editor="1",pages="1",title="0",year="1"}}
     \if@filesw
     \immediate\write\@auxout{\string\citation{apsrev41Control}}%
    \fi
}%

\begin{document}

\title{Collision-resolved pressure sensing}
\author{Daniel S. Barker}\thanks{daniel.barker@nist.gov}
\affiliation{Sensor Science Division, National Institute of Standards and Technology, Gaithersburg, MD}
\author{Daniel Carney}\thanks{carney@lbl.gov}
\affiliation{Physics Division, Lawrence Berkeley National Laboratory, Berkeley, CA}
\author{Thomas W. LeBrun}
\affiliation{Microsystems and Nanotechnology Division, National Institute of Standards and Technology, Gaithersburg, MD}
\author{David C. Moore}
\affiliation{Wright Laboratory, Department of Physics, Yale University, New Haven, CT}
\author{Jacob M. Taylor}
\affiliation{Joint Quantum Institute, University of Maryland, College Park, MD}

\date{\today}

\begin{abstract}
Heat and pressure are ultimately transmitted via quantized degrees of freedom, like gas particles and phonons.
While a continuous Brownian description of these noise sources is adequate to model measurements with relatively long integration times, sufficiently precise measurements can resolve the detailed time dependence coming from individual bath-system interactions.
We propose the use of nanomechanical devices operated with impulse readout sensitivity around the ``standard quantum limit'' to sense ultra-low gas pressures by directly counting the individual collisions of gas particles on a sensor.
We illustrate this in two paradigmatic model systems: an optically levitated nanobead and a tethered membrane system in a phononic bandgap shield.
\end{abstract}

\maketitle

Mechanical objects placed in imperfect vacuum are subject to heat and pressure from their environments. While measurements of the motion of the mechanical object over long timescales will detect these thermal backgrounds as continuous random Brownian motion of the system \cite{brown,einstein}, measurements at very fast timescales can be sensitive to the individual microscopic system-environment interactions \cite{li2010measurement,huang2011direct,franosch2011resonances}, a regime in which the continuous Brownian description breaks down. 

In this paper, we suggest methods to detect gas pressure at this single-quantum limit using mechanical sensors operated at or near the quantum readout regime \cite{blencowe2004quantum,kippenberg2008cavity,aspelmeyer2014cavity}. This would represent pressure sensing at its fundamental limit, relevant in ultra-low pressure environments with small devices. This level of environmental isolation is of increasing importance in a diverse array of contexts, ranging from searches for dark matter \cite{riedel2013direct,Carney:2019pza,carney2021mechanical,afek2022coherent} and other fundamental physics targets \cite{gabrielse1990thousandfold,bose2017spin,carney2019tabletop,Carney:2022pku} to trapped ion quantum computers \cite{pagano2018cryogenic}. In particular, development of pressure sensors capable of operating in extreme high vacuum (XHV, $P \leq 10^{-9}~{\rm Pa}$ \cite{redhead1999extreme}) is an open frontier in precision metrology \cite{scherschligt2017development}.

To estimate the regime where the continuous thermal noise model breaks down, consider a small mechanical element of mass $m_s$ and cross-sectional area $A$ in a dilute ideal gas with pressure $P$ and temperature $T$. The ambient gas particles, with mass $m_g$, collide with the sensor and impart momentum kicks of order $\Delta p_T \approx \sqrt{m_g k_B T}$. These kicks occur at an average rate of order
\be
\label{rate-estimate}
\Gamma = \frac{PA}{\Delta p_T} \approx 3~{\rm Hz} \times \left( \frac{P}{10^{-10}~{\rm Pa}} \right) \left( \frac{A}{0.1~{\rm \mu m}^2} \right).
\ee
In the low pressure, small sensor regime, we see this rate can be on the order of one to 100 collisions per second.  Here, we used the Boltzmann distribution to compute the typical velocity of the gas particles, taken to be diatomic hydrogen $m_g \approx 2~{\rm u}$ at room temperature $T = 300~{\rm K}$. To resolve such a kick, the sensor needs to be operated with sensitivity $\Delta p \lesssim \Delta p_T \approx 7~{\rm keV}/c$ and with a bandwidth $1/\tau >\Gamma$ where $\tau$ is the integration time for a measurement of a single kick.

We now ask: can these weak kicks be resolved by a macroscopic sensor? One simple answer is given by comparing with the standard quantum limit (SQL) for impulses \cite{clerk2004quantum,ghosh2020backaction},
\begin{align}
\label{sql}
\Delta p_{\rm SQL} = \sqrt{\frac{\hbar m_s}{\tau}} \approx 0.8~{\rm keV}/c \times \left( \frac{m_s}{1~{\rm fg}} \right)^{1/2} \left( \frac{1~{\rm ms}}{\tau} \right)^{1/2}.
\end{align}
The sensor mass $m_s$ in this example is benchmarked against a $50~{\rm nm}$ radius silica sphere for comparison with \eqref{rate-estimate}. We note that simple numerical differentiation of a series of position measurements at the position SQL yields the impulse SQL, a limit which has been achieved to good approximation in a number of nanomechanical devices \cite{delic2020cooling,tebbenjohanns2021quantum}. Taken together, these numbers indicate that quantum-limited nanomechanical devices \cite{blencowe2004quantum,kippenberg2008cavity,aspelmeyer2014cavity} in ultra-high (UHV) or extreme-high vacuum (XHV), monitored for impulses at sub-second integration times and with near-SQL sensitivity, could be sensitive to discrete kicks from the ambient gas, as suggested in \cite{ghosh2020backaction,magrini2021real,afek2022coherent}.

In what follows, we provide more detailed calculations and proposals toward achieving such measurements. Our primary concern will be on feasibility of achieving the relevant limits above, especially the bandwidth requirements: the quantum noise \eqref{sql} scales favorably with longer measurement time, but this must be balanced against common technical noise sources with flat power, which lead to $\Delta p_{\rm tech} \sim \sqrt{\tau}$. As practical examples, we study the possible use of levitated optomechanical nanospheres as well as tethered membranes in a phononic bandgap shield as a pair of complementary platforms.

\section{Mechanical impulse sensing}

\begin{figure}[t]
\includegraphics[scale=.5]{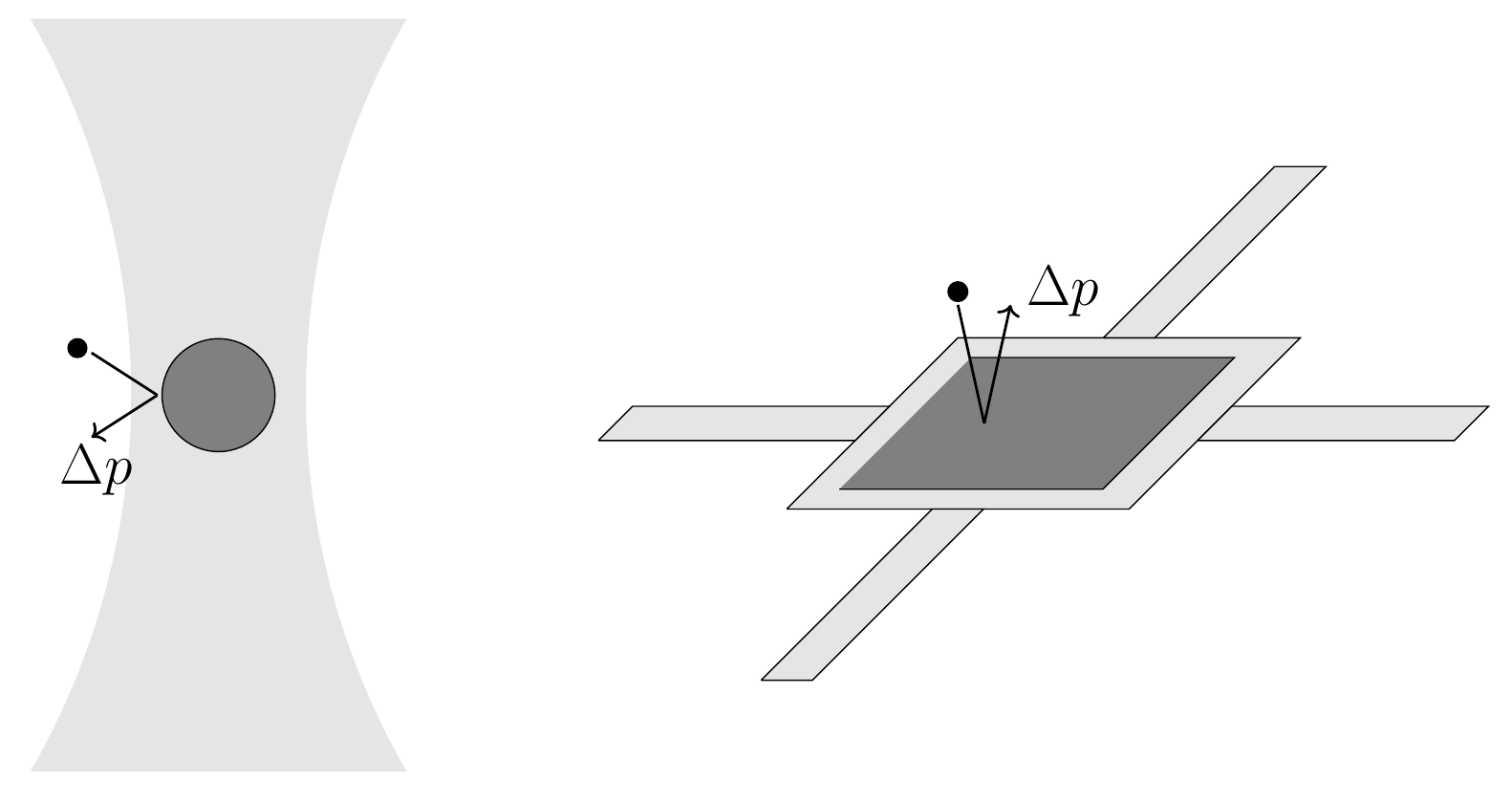}
\caption{Schematics of the basic detection scheme, with either a levitated nanoparticle (left) or tethered membrane in the unit cell of a phononic bandgap shield (right). When an environmental gas particle collides with the mechanical element, it deposits momentum $\Delta p$, which can be detected by continuously monitoring the position $x(t)$ of the element.}
\label{fig-cartoons}
\end{figure}

We will consider opto- or electro-mechanical devices operated as impulse sensors. These devices consist of a mode of a mechanical element of mass $m_s$, which we approximate as executing harmonic motion at frequency $\omega_s$, continuously monitored by an optical or microwave field. Typically one monitors the position $x(t)$ of the mechanics; assuming we have knowledge of the linear response of the device to an input force $x(\nu) = \chi(\nu) F(\nu)$, where $\chi(\nu)$ is a response function in the frequency domain, we can infer the applied force time series $F(t)$. See Fig. \ref{fig-cartoons}. 

First, consider optically monitoring the center-of-mass motion $x(t)$ of a levitated dielectric bead \cite{yin2013optomechanics,millen2020optomechanics,moore2021searching}.
Levitation of dielectric beads with radii ranging from $50~{\rm nm}$ to $10~{\rm \mu m}$ and oscillation frequencies in the 0.1~kHz to 1~MHz range has been demonstrated. In particular, very recently, a pair of experiments have demonstrated feedback cooling to the center-of-mass ground state in optically levitated beads with radius around $100~{\rm nm}$, trapped at around $\omega_s/2\pi \approx 100~{\rm kHz}$ \cite{delic2020cooling,tebbenjohanns2021quantum}. This feedback cooling mechanism operates by continuously monitoring the bead's position fluctuations and applying feedback kicks in order to drive it to the ground state. To reach the ground state this way requires precisely that one can monitor the fluctuations near the SQL, corresponding to the ground state uncertainty $\Delta x_{SQL} = \sqrt{\hbar/m_s \omega_s}$ of the mass. Thus these systems are already operating in the SQL regime, although at two orders of magnitude higher frequency than the optimal integration time assumed in \eqref{sql}. Even at this sensitivity, they should be capable of sensing the high-energy tail of the Boltzmann distribution \cite{magrini2021real}.

Alternatively, one could consider a clamped system like a membrane. In this approach, the center-of-mass of the membrane is fixed, and one monitors the amplitude of the vibrations, for example of the fundamental mode \cite{jayich2008dispersive}. These vibrational modes tend to be of higher frequency than center-of-mass motion, and so obtaining the same SQL sensitivity will require lower-mass devices. For example, a square graphene monolayer around $20~{\rm nm}$ on each side would have a mass around $m_s \approx 10^{-3}~{\rm fg}$, and thus could achieve the required sensitivity if its fundamental mode could be tuned to around $\omega_s/2\pi \approx 1~{\rm MHz}$ \cite{bunch2007electromechanical}. One could also consider membranes constructed from non-conductive materials like silicon nitride \cite{zwickl2008high,wilson2009cavity}.

Either the clamped membrane or levitated bead can be continuously operated as a detector of sharp impulse signals
\be
F_{\rm sig}(t) \approx \Delta p \delta(t - t_0).
\ee
As described above, one monitors the position $x(t)$ as a time series; an impulse will appear as a kick followed by a ring-down in this data stream. Individual collisions can then be resolved if the size $\Delta p$ of these kicks is large compared to the continuous noise acting on the device. We will describe this noise through its power spectral density (PSD), denoted $S_{FF}(\nu)$, which has dimensions of force$^2$ per frequency. To estimate the amplitude of a signal $F_{\rm sig}(t)$ in a given data time series $F(t)$, the strategy that minimizes the estimator variance is to convolve the data with a matched filter $f(t)$, which weights frequencies by signal-to-noise. For an impulse signal described by a flat spectrum versus frequency, $f(\nu) \sim 1/S_{FF}(\nu)$. With this filter, the signal-to-noise ratio of an impulse $\Delta p$ compared to the noise is given by \cite{ghosh2020backaction}
\be
\label{dp}
\frac{S}{N} = \sqrt{\int_{0}^{\infty} d\nu \frac{\Delta p^2}{S_{FF}(\nu)} }.
\ee
In other words, the best sensitivity is achieved by minimizing the integrated noise PSD. The integral is dominated by a bandwidth $\Delta \nu$, which in turns sets the temporal width $\tau \sim 1/\Delta \nu$ of the filter $f(t)$.

\begin{figure}[t]
\includegraphics[scale=0.45]{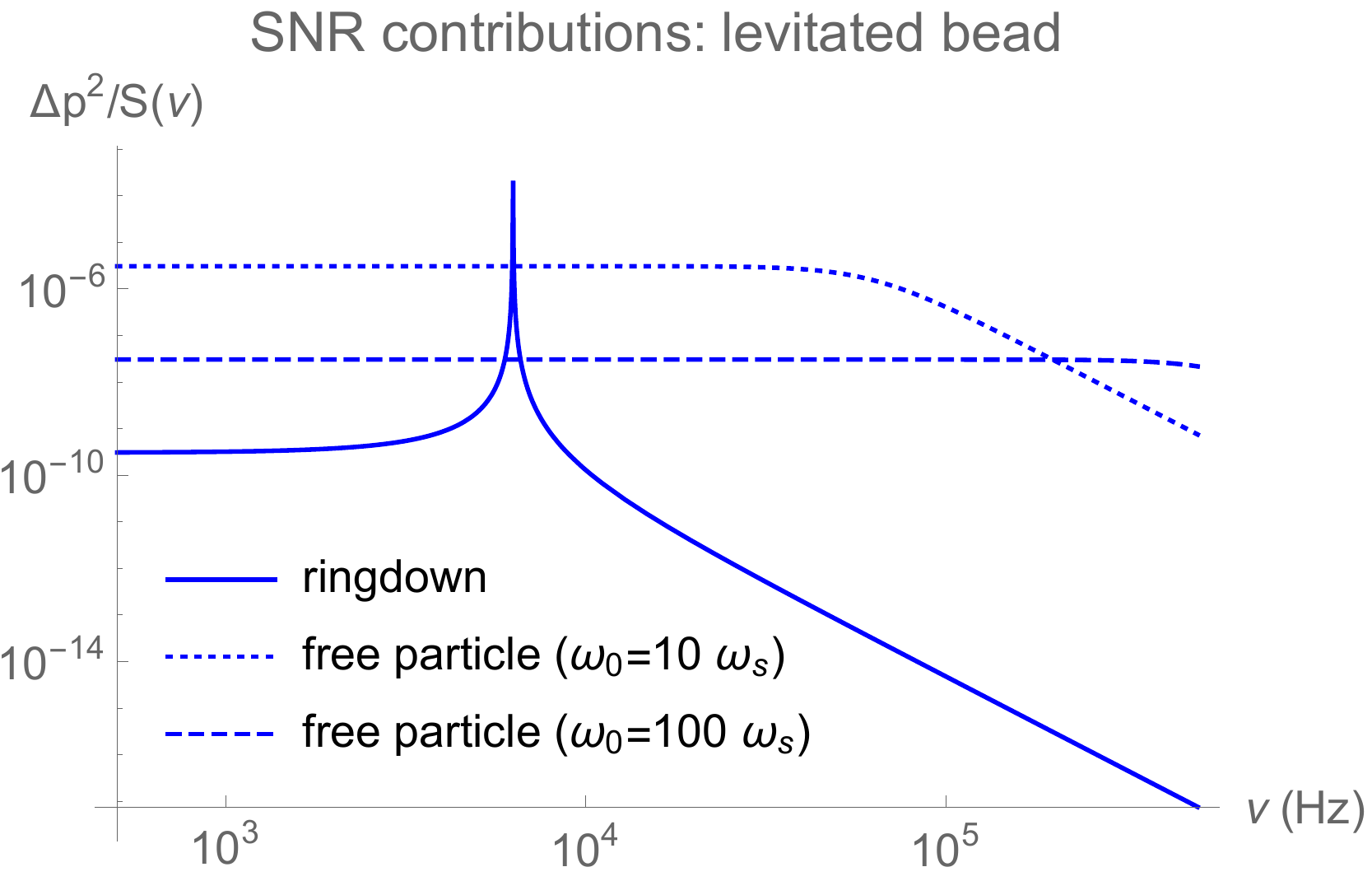}
\caption{Example contributions to the integrand of the total impulse signal-to-noise \eqref{dp}, here shown for a $100~{\rm nm}$ levitated bead, trapped at $\omega_s/2\pi = 1~{\rm kHz}$. The noise is assumed to be dominated only by quantum readout. In the ringdown protocol, the laser power is tuned so that shot noise and backaction are equal at the mechanical resonance $\omega_0 = \omega_s$. In the free particle examples, these noise terms are instead tuned to equality at frequencies $\omega_{0} \gg \omega_s$ above the resonance. Here we use $\Delta p = 7~{\rm keV}/c$, corresponding to collisions with diatomic hydrogen gas at $300~{\rm K}$. The integrated SNR $\approx 2$ for the ringdown case and $\approx 1.5$ for the $\omega_0 = 10 \omega_s$ free particle case.}
\label{fig-snrcontribslev}
\end{figure}

The noise power spectrum of an optomechanical device contains a number of factors with different frequency dependencies. The quantum noise term is what leads to the SQL scaling \eqref{sql}. In general, the quantum noise consists of a term corresponding to shot noise (e.g., phase noise in the readout laser) and a term corresponding to backaction noise (e.g., random radiation pressure exerted on the mechanics by the readout laser). By tuning the readout system appropriately, one can choose a specific fixed frequency $\omega_{0}$ where $S^{\rm shot}_{FF}(\omega_0) = S^{\rm ba}_{FF}(\omega_0)$. We can illustrate this with the example of a levitated free-space optomechanics system, in which \cite{ghosh2020backaction} 
\be
\label{noise-cavity}
S^{\rm Q}_{FF}(\nu) = \hbar |\chi_m(\omega_0)| \left[  \frac{1}{|\chi_m(\nu)|^2} + \frac{1 }{ |\chi_m(\omega_0)|^2} \right]
\ee
in terms of the mechanical response function $\chi_m(\nu) = [m_s (\nu^2 - \omega_s^2 - i \gamma_s \nu)]^{-1}$, where $\gamma_s$ is the damping rate of the mechanics. The usual SQL result \eqref{sql} comes from choosing $\omega_0 = \omega_s$, in which case the noise is sharply minimized on the mechanical resonance $\omega_s$, see the solid curve in Fig. \ref{fig-snrcontribslev}. In this case, one has $S_{FF} \approx 2 \hbar m_s \gamma_s \omega_s$ within a mechanical linewidth $\gamma_s$. Using \eqref{dp}, this means we need a measurement at this narrow bandwidth, i.e. a ringdown measurement with $\tau \sim 1/\gamma_s \gg 1/\omega_s$, to achieve \eqref{sql}.

However, in practice, such a long measurement is not practical. In particular, with a damping rate $\gamma_s \lesssim 1~{\rm Hz}$, multiple gas signals would pile on top of each other [see Eq. \eqref{rate-estimate}]. More fundamentally, in addition to quantum noise, there are technical noises which act as effective heating sources. For example, jittering of the trapping laser in a levitated system or exchange of phonons between a membrane and its support structure will act as approximately white noise sources in a gas collision measurement. These can be approximated as Ohmic heating by a bath with temperature $T_B$, leading to a white noise contribution $S_{FF}^{\rm tech} \sim \gamma k_B T_B$, where $\gamma$ is typically no smaller than the mechanical damping rate. This places a fundamental restriction on achievable bandwidth: they act as a noise with $\Delta p_{\rm tech} = \sqrt{4 m_s k_B T_B \gamma \tau}$, leading to an upper bound on the integration time $\tau$.

There are two strategies to overcome this bandwidth requirement. One is to apply time-dependent cold damping, where the damping coefficient $\gamma_s$ is periodically increased using a noiseless feedback system, so that the signal is distributed over a wider bandwidth (while the signal-to-noise in that bandwidth remains constant), and readout can be performed with a shorter integration time \cite{vitali2001optomechanical,harris2013minimum}. Another is to tune the laser so that $\omega_0 > \omega_s + \gamma_s$. In this case, $S_{Q}(\nu) \approx \hbar m_s \omega_0^2 (1+\nu^4/\omega_{0}^4)$ near $\nu \approx \omega_0$; inserting this into \eqref{dp} then gives \eqref{sql} with $\tau = 1/\omega_0$. See the dashed curves in Fig. \ref{fig-snrcontribslev}. Crucially, however, the integral is here dominated by a band of order $\omega_0$, which corresponds to a much narrower time domain filter. Unlike the ringdown measurement, this protocol essentially treats the mechanical system as a free particle, where the entire impulse and measurement process is faster than a mechanical period. With either the cold damping scheme or above-resonance quantum noise scheme, the key point is that the effective Ohmic heating can be limited to an integration time of order $\tau \sim 1/\omega_0$, much less than the ringdown measurement where $\tau \sim 1/\gamma_s$.

As an important numerical example, consider a tethered membrane system, with which we want to detect gas collisions. Phonons from the clamping substrate can leak into and out of the mechanical element; these will appear as an Ohmic heating background. To see an individual gas collision, we require that the heating from these phonons is subdominant to the collision signal:
\be
\frac{\Delta p_{T}}{\Delta p_{\rm tech}} = \sqrt{\frac{m_g T_{\rm gas}}{m_s T_B} \frac{Q}{\omega_s \tau}} \gtrsim 1,
\ee
where $Q = \gamma/\omega_s$ is the quality factor of the membrane mode. Consider an integration window $\tau \approx 1/\omega_s$, and our nominal $400~{\rm nm}^2$ monolayer device detecting diatomic hydrogen gas. For $T_{\rm gas} = T_B = 300~{\rm K}$, we need $Q \sim 10^{7}$. If the substrate can be made smaller, or held at lower temperatures, the requirements are reduced. With the same mass, but $T_B = 4~{\rm K}$ we need $Q \sim 10^5$; similarly, detecting the $T_{\rm gas} \approx 4~{\rm K}$ helium atoms boiling off the walls of a dilution refrigerator $T_{B} \approx 10~{\rm mK}$ would again require $Q \sim 10^5$. The high $Q$ values required with hot substrates could potentially be obtained with phononic bandgap shielding \cite{yu2014phononic,Kirchhof2021}, as depicted schematically in Fig. \ref{fig-cartoons}. We note also that with sufficiently fast measurements ($\tau \lesssim Q/\hbar k_B T_{B}$) one could try to resolve individual thermal phonons rather than treat them as a continuous background, a task of relevance for example in searches for light dark matter \cite{Knapen:2017ekk}.

\section{Gas collision spectrum}

\begin{figure}[b]
\centering
\includegraphics[scale=.5]{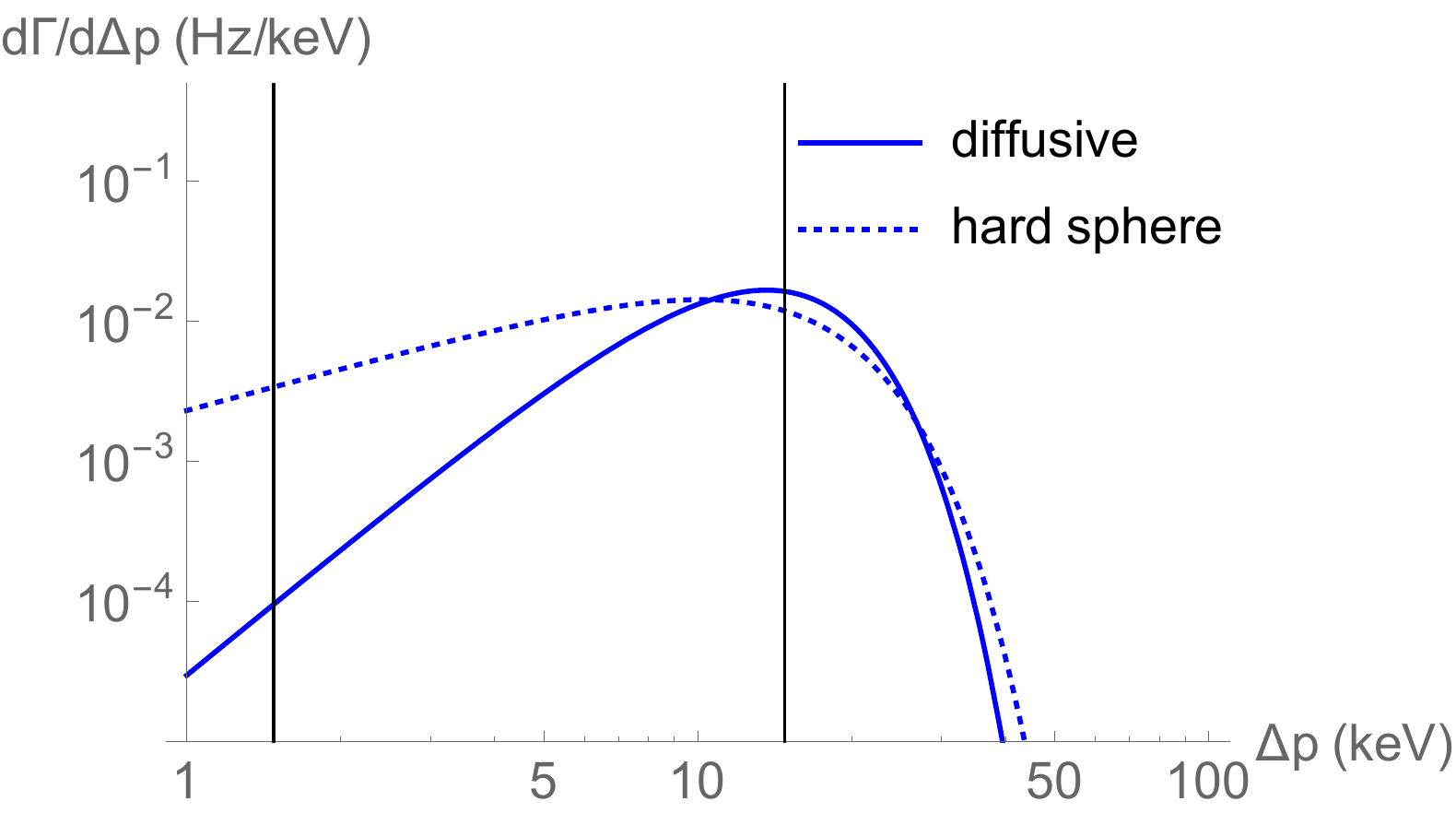}
\caption{Example spectrum of collision events, expressed as a differential rate $d\Gamma$ per given impulse value $\Delta p$. The black lines label the nominal detection threshold $\Delta p_{\rm min} = \Delta p_{\rm SQL}$, with a solid sphere of radius $50~{\rm nm}$, trapped at either $\omega_s/2\pi = 1~{\rm kHz}$ (left) or $100~{\rm kHz}$ (right). We again assume the gas is dominated by diatomic hydrogen at $300~{\rm K}$, and we show the predictions for pure hard sphere scattering as well as diffusive scattering corrections.}
\label{figure-example}
\end{figure}

Collisions of the ambient gas with a mechanical sensor produce a spectrum of impulse signals.
Because the thermal de~Broglie wavelength of H$_2$ at $300$~K is approximately $70$~pm, much smaller than the $10$~nm typical size of an impulse sensor, we can treat the gas-sensor collisions classically.
The background gas can scatter both diffusely and specularly (perfectly reflectively) from the mechanical sensor~\cite{RamsayMolBeams, Cavalleri2010, Martinetz2018, Blakemore2020}.
When the thermal de~Broglie wavelength of the background gas is small compared to the sensor's surface roughness, which will not be the case for atomically flat tethered devices, diffuse scattering will dominate the impulse spectrum.
We therefore include both specular and diffuse scattering to calculate a differential event rate in terms of the momentum transfer in each event:
\be
\label{eq:diffeventrate}
\frac{d\Gamma}{d\Delta p} = \frac{n_g A \Delta p}{4 m_g^2} f_{\rm B}\Big(\frac{\Delta p}{2m_g}\Big)\Big[(1-\alpha) +\alpha\,\xi\Big(\frac{\Delta p}{m_g\overline{v}}\Big)\Big].
\ee
Here, $n_g$ is the number density of the gas with mass $m_g$, $A$ is the total surface area of the sensor, $f_{\rm B}(v)$ is the Boltzmann distribution for velocity $v$ at temperature $T$, and $\overline{v} = \sqrt{k_B T/m_g}$ is the root-mean-square thermal velocity.
The momentum accommodation coefficient $0\le\alpha\le1$ is the fraction of background molecules that scatter diffusely from the sensor.
The factors before the brackets in \eqref{eq:diffeventrate} represent specular reflection; the effect of diffuse reflection is encapsulated by the $\mathcal{O}(1)$ factor
\be
\xi(x) = \sqrt{\pi} x \left( 1 - \frac{2}{x^2} \right) \mathrm{erf}\left( \frac{x}{2} \right) e^{-x^2/8} + 2 e^{-3 x^2/8},
\ee
where $x = \Delta p/m_g \overline{v}$ is a dimensionless measure of the momentum transfer, and $\mathrm{erf}$ is the Gaussian error function. An example is plotted in Fig.~\ref{figure-example}.

The total detectable event rate $\Gamma(\Delta p_{\rm min})$ can be derived from the differential event rate by integrating over $\Delta p$ from our detection threshold $\Delta p_{\rm min}$ to infinity. This gives
\be
\label{eq:totalrate}
\Gamma(\Delta p_{\rm min}) = \frac{n_g A \overline{v}}{\sqrt{2\pi}} \Big[(1-\alpha)\eta_s\Big(\frac{\Delta p_{\rm min}}{m_g\overline{v}}\Big)+\alpha\,\eta_d\Big(\frac{\Delta p_{\rm min}}{m_g\overline{v}}\Big)\Big],
\ee
where the detectable momentum cutoffs for specular scattering $\eta_s$ and diffuse scattering $\eta_d$ are given by
\be
\label{eq:ssc}
    \eta_s(x_{\rm min}) = e^{-x_{\rm min}^2/8}
\ee
and
\be
\label{eq:dsc}
    \eta_d(x_{\rm min}) = e^{-x_{\rm min}^2/2} + \frac{\sqrt{\pi}}{2}  x_{\rm min} \mathrm{erf} \left( \frac{x_{\rm min}}{2} \right) e^{-x_{\rm min}^2/4},
\ee
respectively.
In (\ref{eq:ssc}) and (\ref{eq:dsc}), $x_{\rm min} = \Delta p_{\rm min}/m_g \overline{v}$.
In the $\Delta p_{\rm min} \rightarrow 0$ limit, we have $\eta_s, \,\eta_d \rightarrow 1$, and Eq.~\eqref{eq:totalrate} simplifies to the standard result from kinetic gas theory ($\Gamma = n A \overline{v}/\sqrt{2\pi}$) or scattering theory ($\Gamma = n \langle\sigma v\rangle$, where $\langle \cdots \rangle$ denotes a thermal average).
Our result~\eqref{eq:totalrate} assumes detectability of impulses on all three spatial axes; if one is monitoring only one or two axes there is an additional geometric factor, given in detail in the supplemental material.

\section{Applications}

\subsection{\label{sec:psensor} Primary pressure sensing}

Direct detection of background gas molecules through collision counting opens the possibility of primary pressure sensing in the ultra-high vacuum (UHV, $10^{-9}$~Pa $\le P<10^{-6}$~Pa) and extreme-high vacuum (XHV, $P<10^{-9}$~Pa) regimes using mechanical systems~\cite{scherschligt2017development}.
Prior mechanical vacuum sensors have been based on damping measurements and therefore limited to the high vacuum range (HV, $10^{-6}$~Pa $\le P<10^{-1}$~Pa)~\cite{Fremerey1985, Tilford1992, Scherschligt2018, Blakemore2020}.
We can use the ideal gas law and invert Eq.~(\ref{eq:totalrate}) to find the measurement equation for the pressure of the mechanical collision sensor
\begin{equation}
    \begin{split}
    \label{eq:measeq}
        P = \Gamma(\Delta p_{\rm min}) \frac{\sqrt{2 \pi} k_B T}{A \overline{v} [(1-\alpha)\eta_s(x_{\rm min})+\alpha\eta_d(x_{\rm min})]}.
    \end{split}
\end{equation}
At first, the presence of an accommodation coefficient in Eq.~\eqref{eq:measeq} appears to prevent an optomechanical collision sensor from operating as a primary gauge, since $\alpha$ depends on the surface roughness of the sensor and may vary significantly from sensor to sensor.
However, the accommodation coefficient drops out when the optomechanical system detects all background gas collisions ($\eta_s,\,\eta_d\rightarrow 1$).
Because $\eta_s$ increases more slowly than $\eta_d$, we estimate that $\eta_s > 0.99$, which occurs when $\Delta p_{\rm min} < m_g\overline{v}/4 \approx 1.7~\text{keV}/c$ for H$_2$ at $300$~K, is sufficient for the collision sensor to be primary.
Even when detectable momentum cutoffs are significantly less than one, pseudo-primary operation can be recovered in two ways.
First, collisions of xenon with a nanosphere with $1$~nm surface roughness will be approximately $98\%$ diffuse~\cite{Hsu2018}.
Second, collisions of H$_2$ or helium with an ultraflat tethered 2D material will be greater than $99\%$ specular~\cite{Lui2009}.
In either case, template momentum spectra for other gases can be built up ratiometrically~\cite{scherschligt2017development}.

Eq.~\ref{eq:measeq} contains two constants ($k_B$ and $m_g$) and three measured quantities ($\Gamma(\Delta p_{\rm min})$, $A$, and $T$).
It is therefore traceable to the second, meter, kilogram, and kelvin.
Assuming the measurement of $\Gamma(\Delta p_{\rm min})$ is limited by molecule arrival shot noise, the nanosphere sensor plotted in Fig.~\ref{figure-example} would reach $1\%$ statistical uncertainty approximately 400 times faster than a deployable primary vacuum sensor based on laser-cooled atoms~\cite{Ehinger2022}.
The surface area of a nanosphere can be determined by combining an \textit{in-situ} mass measurement with prior scanning electron microscope characterization~\cite{Blakemore2019}, while the area of a tethered device can be measured during fabrication.
Calibrated or primary contact thermometers can measure the gas temperature and the surface temperature of a tethered sensor~\cite{Purdy2017}.
The surface temperature of a nanosphere can be estimated from the gas temperature by heat transfer modelling or be measured using an infrared thermometer~\cite{Hebestreit2018, Vinante2019}.
We note that thermal equilibrium between the gas and the sensor is not strictly required provided that $\eta_s \approx \eta_d \approx 1$ can be maintained.
We believe that a mechanical collision counter can achieve a total (statistical and non-statistical) pressure measurement uncertainty at the few percent level, which is competitive with cold-atom vacuum standards~\cite{Ehinger2022}.

\subsection{Gas analysis}

Up to this point, we have considered background gases that consist of a single species.
In a real vacuum environment, the background gas will contain a variety of species and the differential event rate becomes
\be
\label{eq:gasanalysis}
\frac{d\Gamma}{d\Delta p} = \mathlarger{\sum}_i\,\frac{n_{g,i} A \Delta p}{4 m_{g,i}^2} f_{\rm B}\Big(\frac{\Delta p}{2m_{g,i}}\Big)\Big[(1-\alpha) +\alpha\,\xi\Big(\frac{\Delta p}{m_{g,i}\overline{v}_i}\Big)\Big],
\ee
where the sum runs over all background gas species $i$.
Because the peak event rate due to gas $i$ occurs roughly at $\Delta p = 2 m_{g,i} \overline{v}_i$, we can use measurements of the differential event rate at several resolvable momenta $\Delta p$ to extract all background gas densities $n_{g,i}$ (or, equivalently, partial pressures $P_i$).
Fully disentangling the overlapping event distributions requires detailed knowledge of the characteristic momentum spectrum of each gas (using the ratiometric method discussed in Sec.~\ref{sec:psensor}) and represents a considerable data analysis challenge.
However, the resulting collision counting gas analyzer has three significant advantages over conventional quadrupole mass spectrometers.
First, it is primary (see Sec.~\ref{sec:psensor}), allowing gas analysis in applications were periodic calibrations are difficult or impossible.
Second, it is intrinsically low outgassing, permitting analysis deep in the XHV where quadrupole spectrometers may add large systematic uncertainty.
Finally, it is chip-scale, so leak detection can be performed in compact, autonomous systems.

\section{Outlook}

In a sufficiently good vacuum, the only way to sense ambient gas pressure is to detect individual gas collisions with a sensor. Here, we outlined two architectures for such detection using mechanical sensors operated in the quantum readout regime. As a practical application, this would enable a pressure standard capable of operation in extreme high vacuum, a little-explored but increasingly important environment. At a fundamental level, such a device would represent the sensing of pressure at its ultimate limit, where the very concept of continuous pressure breaks down, and one requires a description in terms of individual quanta. 

\begin{acknowledgments}
We thank Stephen Eckel, James Fedchak, Sinead Griffin, Lorenzo Magrini, Archana Raja, Cindy Regal, Benjamin Reschovsky, Alp Sipahigil, and Dalziel Wilson for discussions. DC is supported by the US Department of Energy under contract DE-AC02-05CH11231 and Quantum Information Science Enabled Discovery (QuantISED) for High Energy Physics grant KA2401032.
\end{acknowledgments}

\bibliography{gas_sensing.bib}

\begin{thebibliography}{51}%
\makeatletter
\providecommand \@ifxundefined [1]{%
 \@ifx{#1\undefined}
}%
\providecommand \@ifnum [1]{%
 \ifnum #1\expandafter \@firstoftwo
 \else \expandafter \@secondoftwo
 \fi
}%
\providecommand \@ifx [1]{%
 \ifx #1\expandafter \@firstoftwo
 \else \expandafter \@secondoftwo
 \fi
}%
\providecommand \natexlab [1]{#1}%
\providecommand \enquote  [1]{``#1''}%
\providecommand \bibnamefont  [1]{#1}%
\providecommand \bibfnamefont [1]{#1}%
\providecommand \citenamefont [1]{#1}%
\providecommand \href@noop [0]{\@secondoftwo}%
\providecommand \href [0]{\begingroup \@sanitize@url \@href}%
\providecommand \@href[1]{\@@startlink{#1}\@@href}%
\providecommand \@@href[1]{\endgroup#1\@@endlink}%
\providecommand \@sanitize@url [0]{\catcode `\\12\catcode `\$12\catcode
  `\&12\catcode `\#12\catcode `\^12\catcode `\_12\catcode `\%12\relax}%
\providecommand \@@startlink[1]{}%
\providecommand \@@endlink[0]{}%
\providecommand \url  [0]{\begingroup\@sanitize@url \@url }%
\providecommand \@url [1]{\endgroup\@href {#1}{\urlprefix }}%
\providecommand \urlprefix  [0]{URL }%
\providecommand \Eprint [0]{\href }%
\providecommand \doibase [0]{http://dx.doi.org/}%
\providecommand \selectlanguage [0]{\@gobble}%
\providecommand \bibinfo  [0]{\@secondoftwo}%
\providecommand \bibfield  [0]{\@secondoftwo}%
\providecommand \translation [1]{[#1]}%
\providecommand \BibitemOpen [0]{}%
\providecommand \bibitemStop [0]{}%
\providecommand \bibitemNoStop [0]{.\EOS\space}%
\providecommand \EOS [0]{\spacefactor3000\relax}%
\providecommand \BibitemShut  [1]{\csname bibitem#1\endcsname}%
\let\auto@bib@innerbib\@empty
\bibitem [{\citenamefont {Brown}(1828)}]{brown}%
  \BibitemOpen
  \bibfield  {author} {\bibinfo {author} {\bibfnamefont {R.}~\bibnamefont
  {Brown}},\ }\bibfield  {title} {\enquote {\bibinfo {title} {A brief account
  of microscopical observations made in the months of june, july and august
  1827, on the particles contained in the pollen of plants; and on the general
  existence of active molecules in organic and inorganic bodies},}\ }\href
  {\doibase 10.1080/14786442808674769} {\bibfield  {journal} {\bibinfo
  {journal} {The Philosophical Magazine}\ }\textbf {\bibinfo {volume} {4}},\
  \bibinfo {pages} {161--173} (\bibinfo {year} {1828})}\BibitemShut {NoStop}%
\bibitem [{\citenamefont {Einstein}(1905)}]{einstein}%
  \BibitemOpen
  \bibfield  {author} {\bibinfo {author} {\bibfnamefont {A.}~\bibnamefont
  {Einstein}},\ }\bibfield  {title} {\enquote {\bibinfo {title} {Über die von
  der molekularkinetischen theorie der wärme geforderte bewegung von in
  ruhenden flüssigkeiten suspendierten teilchen},}\ }\href {\doibase
  https://doi.org/10.1002/andp.19053220806} {\bibfield  {journal} {\bibinfo
  {journal} {Annalen der Physik}\ }\textbf {\bibinfo {volume} {322}},\ \bibinfo
  {pages} {549--560} (\bibinfo {year} {1905})}\BibitemShut {NoStop}%
\bibitem [{\citenamefont {Li}\ \emph {et~al.}(2010)\citenamefont {Li},
  \citenamefont {Kheifets}, \citenamefont {Medellin},\ and\ \citenamefont
  {Raizen}}]{li2010measurement}%
  \BibitemOpen
  \bibfield  {author} {\bibinfo {author} {\bibfnamefont {T.}~\bibnamefont
  {Li}}, \bibinfo {author} {\bibfnamefont {S.}~\bibnamefont {Kheifets}},
  \bibinfo {author} {\bibfnamefont {D.}~\bibnamefont {Medellin}}, \ and\
  \bibinfo {author} {\bibfnamefont {M.~G.}\ \bibnamefont {Raizen}},\ }\bibfield
   {title} {\enquote {\bibinfo {title} {Measurement of the instantaneous
  velocity of a brownian particle},}\ }\href@noop {} {\bibfield  {journal}
  {\bibinfo  {journal} {Science}\ }\textbf {\bibinfo {volume} {328}},\ \bibinfo
  {pages} {1673--1675} (\bibinfo {year} {2010})}\BibitemShut {NoStop}%
\bibitem [{\citenamefont {Huang}\ \emph {et~al.}(2011)\citenamefont {Huang},
  \citenamefont {Chavez}, \citenamefont {Taute}, \citenamefont {Luki{\'c}},
  \citenamefont {Jeney}, \citenamefont {Raizen},\ and\ \citenamefont
  {Florin}}]{huang2011direct}%
  \BibitemOpen
  \bibfield  {author} {\bibinfo {author} {\bibfnamefont {R.}~\bibnamefont
  {Huang}}, \bibinfo {author} {\bibfnamefont {I.}~\bibnamefont {Chavez}},
  \bibinfo {author} {\bibfnamefont {K.~M.}\ \bibnamefont {Taute}}, \bibinfo
  {author} {\bibfnamefont {B.}~\bibnamefont {Luki{\'c}}}, \bibinfo {author}
  {\bibfnamefont {S.}~\bibnamefont {Jeney}}, \bibinfo {author} {\bibfnamefont
  {M.~G.}\ \bibnamefont {Raizen}}, \ and\ \bibinfo {author} {\bibfnamefont
  {E.-L.}\ \bibnamefont {Florin}},\ }\bibfield  {title} {\enquote {\bibinfo
  {title} {Direct observation of the full transition from ballistic to
  diffusive brownian motion in a liquid},}\ }\href@noop {} {\bibfield
  {journal} {\bibinfo  {journal} {Nature Physics}\ }\textbf {\bibinfo {volume}
  {7}},\ \bibinfo {pages} {576--580} (\bibinfo {year} {2011})}\BibitemShut
  {NoStop}%
\bibitem [{\citenamefont {Franosch}\ \emph {et~al.}(2011)\citenamefont
  {Franosch}, \citenamefont {Grimm}, \citenamefont {Belushkin}, \citenamefont
  {Mor}, \citenamefont {Foffi}, \citenamefont {Forr{\'o}},\ and\ \citenamefont
  {Jeney}}]{franosch2011resonances}%
  \BibitemOpen
  \bibfield  {author} {\bibinfo {author} {\bibfnamefont {T.}~\bibnamefont
  {Franosch}}, \bibinfo {author} {\bibfnamefont {M.}~\bibnamefont {Grimm}},
  \bibinfo {author} {\bibfnamefont {M.}~\bibnamefont {Belushkin}}, \bibinfo
  {author} {\bibfnamefont {F.~M.}\ \bibnamefont {Mor}}, \bibinfo {author}
  {\bibfnamefont {G.}~\bibnamefont {Foffi}}, \bibinfo {author} {\bibfnamefont
  {L.}~\bibnamefont {Forr{\'o}}}, \ and\ \bibinfo {author} {\bibfnamefont
  {S.}~\bibnamefont {Jeney}},\ }\bibfield  {title} {\enquote {\bibinfo {title}
  {Resonances arising from hydrodynamic memory in brownian motion},}\
  }\href@noop {} {\bibfield  {journal} {\bibinfo  {journal} {Nature}\ }\textbf
  {\bibinfo {volume} {478}},\ \bibinfo {pages} {85--88} (\bibinfo {year}
  {2011})}\BibitemShut {NoStop}%
\bibitem [{\citenamefont {Blencowe}(2004)}]{blencowe2004quantum}%
  \BibitemOpen
  \bibfield  {author} {\bibinfo {author} {\bibfnamefont {M.}~\bibnamefont
  {Blencowe}},\ }\bibfield  {title} {\enquote {\bibinfo {title} {Quantum
  electromechanical systems},}\ }\href@noop {} {\bibfield  {journal} {\bibinfo
  {journal} {Physics Reports}\ }\textbf {\bibinfo {volume} {395}},\ \bibinfo
  {pages} {159--222} (\bibinfo {year} {2004})}\BibitemShut {NoStop}%
\bibitem [{\citenamefont {Kippenberg}\ and\ \citenamefont
  {Vahala}(2008)}]{kippenberg2008cavity}%
  \BibitemOpen
  \bibfield  {author} {\bibinfo {author} {\bibfnamefont {T.~J.}\ \bibnamefont
  {Kippenberg}}\ and\ \bibinfo {author} {\bibfnamefont {K.~J.}\ \bibnamefont
  {Vahala}},\ }\bibfield  {title} {\enquote {\bibinfo {title} {Cavity
  optomechanics: back-action at the mesoscale},}\ }\href@noop {} {\bibfield
  {journal} {\bibinfo  {journal} {Science}\ }\textbf {\bibinfo {volume}
  {321}},\ \bibinfo {pages} {1172--1176} (\bibinfo {year} {2008})}\BibitemShut
  {NoStop}%
\bibitem [{\citenamefont {Aspelmeyer}\ \emph {et~al.}(2014)\citenamefont
  {Aspelmeyer}, \citenamefont {Kippenberg},\ and\ \citenamefont
  {Marquardt}}]{aspelmeyer2014cavity}%
  \BibitemOpen
  \bibfield  {author} {\bibinfo {author} {\bibfnamefont {M.}~\bibnamefont
  {Aspelmeyer}}, \bibinfo {author} {\bibfnamefont {T.~J.}\ \bibnamefont
  {Kippenberg}}, \ and\ \bibinfo {author} {\bibfnamefont {F.}~\bibnamefont
  {Marquardt}},\ }\bibfield  {title} {\enquote {\bibinfo {title} {Cavity
  optomechanics},}\ }\href@noop {} {\bibfield  {journal} {\bibinfo  {journal}
  {Reviews of Modern Physics}\ }\textbf {\bibinfo {volume} {86}},\ \bibinfo
  {pages} {1391} (\bibinfo {year} {2014})}\BibitemShut {NoStop}%
\bibitem [{\citenamefont {Riedel}(2013)}]{riedel2013direct}%
  \BibitemOpen
  \bibfield  {author} {\bibinfo {author} {\bibfnamefont {C.~J.}\ \bibnamefont
  {Riedel}},\ }\bibfield  {title} {\enquote {\bibinfo {title} {Direct detection
  of classically undetectable dark matter through quantum decoherence},}\
  }\href@noop {} {\bibfield  {journal} {\bibinfo  {journal} {Physical Review
  D}\ }\textbf {\bibinfo {volume} {88}},\ \bibinfo {pages} {116005} (\bibinfo
  {year} {2013})}\BibitemShut {NoStop}%
\bibitem [{\citenamefont {Carney}\ \emph {et~al.}(2020)\citenamefont {Carney},
  \citenamefont {Ghosh}, \citenamefont {Krnjaic},\ and\ \citenamefont
  {Taylor}}]{Carney:2019pza}%
  \BibitemOpen
  \bibfield  {author} {\bibinfo {author} {\bibfnamefont {D.}~\bibnamefont
  {Carney}}, \bibinfo {author} {\bibfnamefont {S.}~\bibnamefont {Ghosh}},
  \bibinfo {author} {\bibfnamefont {G.}~\bibnamefont {Krnjaic}}, \ and\
  \bibinfo {author} {\bibfnamefont {J.~M.}\ \bibnamefont {Taylor}},\ }\bibfield
   {title} {\enquote {\bibinfo {title} {{Proposal for gravitational direct
  detection of dark matter}},}\ }\href {\doibase 10.1103/PhysRevD.102.072003}
  {\bibfield  {journal} {\bibinfo  {journal} {Phys. Rev. D}\ }\textbf {\bibinfo
  {volume} {102}},\ \bibinfo {pages} {072003} (\bibinfo {year} {2020})},\
  \Eprint {http://arxiv.org/abs/1903.00492} {arXiv:1903.00492 [hep-ph]}
  \BibitemShut {NoStop}%
\bibitem [{\citenamefont {Carney}\ \emph {et~al.}(2021)\citenamefont {Carney},
  \citenamefont {Krnjaic}, \citenamefont {Moore}, \citenamefont {Regal},
  \citenamefont {Afek}, \citenamefont {Bhave}, \citenamefont {Brubaker},
  \citenamefont {Corbitt}, \citenamefont {Cripe}, \citenamefont {Crisosto}
  \emph {et~al.}}]{carney2021mechanical}%
  \BibitemOpen
  \bibfield  {author} {\bibinfo {author} {\bibfnamefont {D.}~\bibnamefont
  {Carney}}, \bibinfo {author} {\bibfnamefont {G.}~\bibnamefont {Krnjaic}},
  \bibinfo {author} {\bibfnamefont {D.~C.}\ \bibnamefont {Moore}}, \bibinfo
  {author} {\bibfnamefont {C.~A.}\ \bibnamefont {Regal}}, \bibinfo {author}
  {\bibfnamefont {G.}~\bibnamefont {Afek}}, \bibinfo {author} {\bibfnamefont
  {S.}~\bibnamefont {Bhave}}, \bibinfo {author} {\bibfnamefont
  {B.}~\bibnamefont {Brubaker}}, \bibinfo {author} {\bibfnamefont
  {T.}~\bibnamefont {Corbitt}}, \bibinfo {author} {\bibfnamefont
  {J.}~\bibnamefont {Cripe}}, \bibinfo {author} {\bibfnamefont
  {N.}~\bibnamefont {Crisosto}},  \emph {et~al.},\ }\bibfield  {title}
  {\enquote {\bibinfo {title} {Mechanical quantum sensing in the search for
  dark matter},}\ }\href@noop {} {\bibfield  {journal} {\bibinfo  {journal}
  {Quantum Science and Technology}\ }\textbf {\bibinfo {volume} {6}},\ \bibinfo
  {pages} {024002} (\bibinfo {year} {2021})}\BibitemShut {NoStop}%
\bibitem [{\citenamefont {Afek}\ \emph {et~al.}(2022)\citenamefont {Afek},
  \citenamefont {Carney},\ and\ \citenamefont {Moore}}]{afek2022coherent}%
  \BibitemOpen
  \bibfield  {author} {\bibinfo {author} {\bibfnamefont {G.}~\bibnamefont
  {Afek}}, \bibinfo {author} {\bibfnamefont {D.}~\bibnamefont {Carney}}, \ and\
  \bibinfo {author} {\bibfnamefont {D.~C.}\ \bibnamefont {Moore}},\ }\bibfield
  {title} {\enquote {\bibinfo {title} {Coherent scattering of low mass dark
  matter from optically trapped sensors},}\ }\href@noop {} {\bibfield
  {journal} {\bibinfo  {journal} {Physical Review Letters}\ }\textbf {\bibinfo
  {volume} {128}},\ \bibinfo {pages} {101301} (\bibinfo {year}
  {2022})}\BibitemShut {NoStop}%
\bibitem [{\citenamefont {Gabrielse}\ \emph {et~al.}(1990)\citenamefont
  {Gabrielse}, \citenamefont {Fei}, \citenamefont {Orozco}, \citenamefont
  {Tjoelker}, \citenamefont {Haas}, \citenamefont {Kalinowsky}, \citenamefont
  {Trainor},\ and\ \citenamefont {Kells}}]{gabrielse1990thousandfold}%
  \BibitemOpen
  \bibfield  {author} {\bibinfo {author} {\bibfnamefont {G.}~\bibnamefont
  {Gabrielse}}, \bibinfo {author} {\bibfnamefont {X.}~\bibnamefont {Fei}},
  \bibinfo {author} {\bibfnamefont {L.}~\bibnamefont {Orozco}}, \bibinfo
  {author} {\bibfnamefont {R.}~\bibnamefont {Tjoelker}}, \bibinfo {author}
  {\bibfnamefont {J.}~\bibnamefont {Haas}}, \bibinfo {author} {\bibfnamefont
  {H.}~\bibnamefont {Kalinowsky}}, \bibinfo {author} {\bibfnamefont
  {T.}~\bibnamefont {Trainor}}, \ and\ \bibinfo {author} {\bibfnamefont
  {W.}~\bibnamefont {Kells}},\ }\bibfield  {title} {\enquote {\bibinfo {title}
  {Thousandfold improvement in the measured antiproton mass},}\ }\href@noop {}
  {\bibfield  {journal} {\bibinfo  {journal} {Physical review letters}\
  }\textbf {\bibinfo {volume} {65}},\ \bibinfo {pages} {1317} (\bibinfo {year}
  {1990})}\BibitemShut {NoStop}%
\bibitem [{\citenamefont {Bose}\ \emph {et~al.}(2017)\citenamefont {Bose},
  \citenamefont {Mazumdar}, \citenamefont {Morley}, \citenamefont {Ulbricht},
  \citenamefont {Toro{\v{s}}}, \citenamefont {Paternostro}, \citenamefont
  {Geraci}, \citenamefont {Barker}, \citenamefont {Kim},\ and\ \citenamefont
  {Milburn}}]{bose2017spin}%
  \BibitemOpen
  \bibfield  {author} {\bibinfo {author} {\bibfnamefont {S.}~\bibnamefont
  {Bose}}, \bibinfo {author} {\bibfnamefont {A.}~\bibnamefont {Mazumdar}},
  \bibinfo {author} {\bibfnamefont {G.~W.}\ \bibnamefont {Morley}}, \bibinfo
  {author} {\bibfnamefont {H.}~\bibnamefont {Ulbricht}}, \bibinfo {author}
  {\bibfnamefont {M.}~\bibnamefont {Toro{\v{s}}}}, \bibinfo {author}
  {\bibfnamefont {M.}~\bibnamefont {Paternostro}}, \bibinfo {author}
  {\bibfnamefont {A.~A.}\ \bibnamefont {Geraci}}, \bibinfo {author}
  {\bibfnamefont {P.~F.}\ \bibnamefont {Barker}}, \bibinfo {author}
  {\bibfnamefont {M.}~\bibnamefont {Kim}}, \ and\ \bibinfo {author}
  {\bibfnamefont {G.}~\bibnamefont {Milburn}},\ }\bibfield  {title} {\enquote
  {\bibinfo {title} {Spin entanglement witness for quantum gravity},}\
  }\href@noop {} {\bibfield  {journal} {\bibinfo  {journal} {Physical Review
  Letters}\ }\textbf {\bibinfo {volume} {119}},\ \bibinfo {pages} {240401}
  (\bibinfo {year} {2017})}\BibitemShut {NoStop}%
\bibitem [{\citenamefont {Carney}\ \emph {et~al.}(2019)\citenamefont {Carney},
  \citenamefont {Stamp},\ and\ \citenamefont {Taylor}}]{carney2019tabletop}%
  \BibitemOpen
  \bibfield  {author} {\bibinfo {author} {\bibfnamefont {D.}~\bibnamefont
  {Carney}}, \bibinfo {author} {\bibfnamefont {P.~C.~E.}\ \bibnamefont
  {Stamp}}, \ and\ \bibinfo {author} {\bibfnamefont {J.~M.}\ \bibnamefont
  {Taylor}},\ }\bibfield  {title} {\enquote {\bibinfo {title} {Tabletop
  experiments for quantum gravity: a user’s manual},}\ }\href@noop {}
  {\bibfield  {journal} {\bibinfo  {journal} {Classical and Quantum Gravity}\
  }\textbf {\bibinfo {volume} {36}},\ \bibinfo {pages} {034001} (\bibinfo
  {year} {2019})}\BibitemShut {NoStop}%
\bibitem [{\citenamefont {Carney}\ \emph {et~al.}(2022)\citenamefont {Carney},
  \citenamefont {Leach},\ and\ \citenamefont {Moore}}]{Carney:2022pku}%
  \BibitemOpen
  \bibfield  {author} {\bibinfo {author} {\bibfnamefont {D.}~\bibnamefont
  {Carney}}, \bibinfo {author} {\bibfnamefont {K.~G.}\ \bibnamefont {Leach}}, \
  and\ \bibinfo {author} {\bibfnamefont {D.~C.}\ \bibnamefont {Moore}},\
  }\bibfield  {title} {\enquote {\bibinfo {title} {{Searches for massive
  neutrinos with mechanical quantum sensors}},}\ }\href@noop {} {\  (\bibinfo
  {year} {2022})},\ \Eprint {http://arxiv.org/abs/2207.05883} {arXiv:2207.05883
  [hep-ex]} \BibitemShut {NoStop}%
\bibitem [{\citenamefont {Pagano}\ \emph {et~al.}(2018)\citenamefont {Pagano},
  \citenamefont {Hess}, \citenamefont {Kaplan}, \citenamefont {Tan},
  \citenamefont {Richerme}, \citenamefont {Becker}, \citenamefont
  {Kyprianidis}, \citenamefont {Zhang}, \citenamefont {Birckelbaw},
  \citenamefont {Hernandez} \emph {et~al.}}]{pagano2018cryogenic}%
  \BibitemOpen
  \bibfield  {author} {\bibinfo {author} {\bibfnamefont {G.}~\bibnamefont
  {Pagano}}, \bibinfo {author} {\bibfnamefont {P.}~\bibnamefont {Hess}},
  \bibinfo {author} {\bibfnamefont {H.}~\bibnamefont {Kaplan}}, \bibinfo
  {author} {\bibfnamefont {W.}~\bibnamefont {Tan}}, \bibinfo {author}
  {\bibfnamefont {P.}~\bibnamefont {Richerme}}, \bibinfo {author}
  {\bibfnamefont {P.}~\bibnamefont {Becker}}, \bibinfo {author} {\bibfnamefont
  {A.}~\bibnamefont {Kyprianidis}}, \bibinfo {author} {\bibfnamefont
  {J.}~\bibnamefont {Zhang}}, \bibinfo {author} {\bibfnamefont
  {E.}~\bibnamefont {Birckelbaw}}, \bibinfo {author} {\bibfnamefont
  {M.}~\bibnamefont {Hernandez}},  \emph {et~al.},\ }\bibfield  {title}
  {\enquote {\bibinfo {title} {Cryogenic trapped-ion system for large scale
  quantum simulation},}\ }\href@noop {} {\bibfield  {journal} {\bibinfo
  {journal} {Quantum Science and Technology}\ }\textbf {\bibinfo {volume}
  {4}},\ \bibinfo {pages} {014004} (\bibinfo {year} {2018})}\BibitemShut
  {NoStop}%
\bibitem [{\citenamefont {Redhead}(1999)}]{redhead1999extreme}%
  \BibitemOpen
  \bibfield  {author} {\bibinfo {author} {\bibfnamefont {P.~A.}\ \bibnamefont
  {Redhead}},\ }\href@noop {} {\emph {\bibinfo {title} {Extreme high
  vacuum}}},\ \bibinfo {type} {Tech. Rep.}\ (\bibinfo  {institution} {CERN},\
  \bibinfo {year} {1999})\BibitemShut {NoStop}%
\bibitem [{\citenamefont {Scherschligt}\ \emph {et~al.}(2017)\citenamefont
  {Scherschligt}, \citenamefont {Fedchak}, \citenamefont {Barker},
  \citenamefont {Eckel}, \citenamefont {Klimov}, \citenamefont {Makrides},\
  and\ \citenamefont {Tiesinga}}]{scherschligt2017development}%
  \BibitemOpen
  \bibfield  {author} {\bibinfo {author} {\bibfnamefont {J.}~\bibnamefont
  {Scherschligt}}, \bibinfo {author} {\bibfnamefont {J.~A.}\ \bibnamefont
  {Fedchak}}, \bibinfo {author} {\bibfnamefont {D.~S.}\ \bibnamefont {Barker}},
  \bibinfo {author} {\bibfnamefont {S.}~\bibnamefont {Eckel}}, \bibinfo
  {author} {\bibfnamefont {N.}~\bibnamefont {Klimov}}, \bibinfo {author}
  {\bibfnamefont {C.}~\bibnamefont {Makrides}}, \ and\ \bibinfo {author}
  {\bibfnamefont {E.}~\bibnamefont {Tiesinga}},\ }\bibfield  {title} {\enquote
  {\bibinfo {title} {Development of a new {UHV/XHV} pressure standard (cold
  atom vacuum standard)},}\ }\href@noop {} {\bibfield  {journal} {\bibinfo
  {journal} {Metrologia}\ }\textbf {\bibinfo {volume} {54}},\ \bibinfo {pages}
  {S125} (\bibinfo {year} {2017})}\BibitemShut {NoStop}%
\bibitem [{\citenamefont {Clerk}(2004)}]{clerk2004quantum}%
  \BibitemOpen
  \bibfield  {author} {\bibinfo {author} {\bibfnamefont {A.}~\bibnamefont
  {Clerk}},\ }\bibfield  {title} {\enquote {\bibinfo {title} {Quantum-limited
  position detection and amplification: A linear response perspective},}\
  }\href@noop {} {\bibfield  {journal} {\bibinfo  {journal} {Physical Review
  B}\ }\textbf {\bibinfo {volume} {70}},\ \bibinfo {pages} {245306} (\bibinfo
  {year} {2004})}\BibitemShut {NoStop}%
\bibitem [{\citenamefont {Ghosh}\ \emph {et~al.}(2020)\citenamefont {Ghosh},
  \citenamefont {Carney}, \citenamefont {Shawhan},\ and\ \citenamefont
  {Taylor}}]{ghosh2020backaction}%
  \BibitemOpen
  \bibfield  {author} {\bibinfo {author} {\bibfnamefont {S.}~\bibnamefont
  {Ghosh}}, \bibinfo {author} {\bibfnamefont {D.}~\bibnamefont {Carney}},
  \bibinfo {author} {\bibfnamefont {P.}~\bibnamefont {Shawhan}}, \ and\
  \bibinfo {author} {\bibfnamefont {J.~M.}\ \bibnamefont {Taylor}},\ }\bibfield
   {title} {\enquote {\bibinfo {title} {Backaction-evading impulse measurement
  with mechanical quantum sensors},}\ }\href@noop {} {\bibfield  {journal}
  {\bibinfo  {journal} {Physical Review A}\ }\textbf {\bibinfo {volume}
  {102}},\ \bibinfo {pages} {023525} (\bibinfo {year} {2020})}\BibitemShut
  {NoStop}%
\bibitem [{\citenamefont {Deli{\'c}}\ \emph {et~al.}(2020)\citenamefont
  {Deli{\'c}}, \citenamefont {Reisenbauer}, \citenamefont {Dare}, \citenamefont
  {Grass}, \citenamefont {Vuleti{\'c}}, \citenamefont {Kiesel},\ and\
  \citenamefont {Aspelmeyer}}]{delic2020cooling}%
  \BibitemOpen
  \bibfield  {author} {\bibinfo {author} {\bibfnamefont {U.}~\bibnamefont
  {Deli{\'c}}}, \bibinfo {author} {\bibfnamefont {M.}~\bibnamefont
  {Reisenbauer}}, \bibinfo {author} {\bibfnamefont {K.}~\bibnamefont {Dare}},
  \bibinfo {author} {\bibfnamefont {D.}~\bibnamefont {Grass}}, \bibinfo
  {author} {\bibfnamefont {V.}~\bibnamefont {Vuleti{\'c}}}, \bibinfo {author}
  {\bibfnamefont {N.}~\bibnamefont {Kiesel}}, \ and\ \bibinfo {author}
  {\bibfnamefont {M.}~\bibnamefont {Aspelmeyer}},\ }\bibfield  {title}
  {\enquote {\bibinfo {title} {Cooling of a levitated nanoparticle to the
  motional quantum ground state},}\ }\href@noop {} {\bibfield  {journal}
  {\bibinfo  {journal} {Science}\ }\textbf {\bibinfo {volume} {367}},\ \bibinfo
  {pages} {892--895} (\bibinfo {year} {2020})}\BibitemShut {NoStop}%
\bibitem [{\citenamefont {Tebbenjohanns}\ \emph {et~al.}(2021)\citenamefont
  {Tebbenjohanns}, \citenamefont {Mattana}, \citenamefont {Rossi},
  \citenamefont {Frimmer},\ and\ \citenamefont
  {Novotny}}]{tebbenjohanns2021quantum}%
  \BibitemOpen
  \bibfield  {author} {\bibinfo {author} {\bibfnamefont {F.}~\bibnamefont
  {Tebbenjohanns}}, \bibinfo {author} {\bibfnamefont {M.~L.}\ \bibnamefont
  {Mattana}}, \bibinfo {author} {\bibfnamefont {M.}~\bibnamefont {Rossi}},
  \bibinfo {author} {\bibfnamefont {M.}~\bibnamefont {Frimmer}}, \ and\
  \bibinfo {author} {\bibfnamefont {L.}~\bibnamefont {Novotny}},\ }\bibfield
  {title} {\enquote {\bibinfo {title} {Quantum control of a nanoparticle
  optically levitated in cryogenic free space},}\ }\href@noop {} {\bibfield
  {journal} {\bibinfo  {journal} {Nature}\ }\textbf {\bibinfo {volume} {595}},\
  \bibinfo {pages} {378--382} (\bibinfo {year} {2021})}\BibitemShut {NoStop}%
\bibitem [{\citenamefont {Magrini}\ \emph {et~al.}(2021)\citenamefont
  {Magrini}, \citenamefont {Rosenzweig}, \citenamefont {Bach}, \citenamefont
  {Deutschmann-Olek}, \citenamefont {Hofer}, \citenamefont {Hong},
  \citenamefont {Kiesel}, \citenamefont {Kugi},\ and\ \citenamefont
  {Aspelmeyer}}]{magrini2021real}%
  \BibitemOpen
  \bibfield  {author} {\bibinfo {author} {\bibfnamefont {L.}~\bibnamefont
  {Magrini}}, \bibinfo {author} {\bibfnamefont {P.}~\bibnamefont {Rosenzweig}},
  \bibinfo {author} {\bibfnamefont {C.}~\bibnamefont {Bach}}, \bibinfo {author}
  {\bibfnamefont {A.}~\bibnamefont {Deutschmann-Olek}}, \bibinfo {author}
  {\bibfnamefont {S.~G.}\ \bibnamefont {Hofer}}, \bibinfo {author}
  {\bibfnamefont {S.}~\bibnamefont {Hong}}, \bibinfo {author} {\bibfnamefont
  {N.}~\bibnamefont {Kiesel}}, \bibinfo {author} {\bibfnamefont
  {A.}~\bibnamefont {Kugi}}, \ and\ \bibinfo {author} {\bibfnamefont
  {M.}~\bibnamefont {Aspelmeyer}},\ }\bibfield  {title} {\enquote {\bibinfo
  {title} {Real-time optimal quantum control of mechanical motion at room
  temperature},}\ }\href@noop {} {\bibfield  {journal} {\bibinfo  {journal}
  {Nature}\ }\textbf {\bibinfo {volume} {595}},\ \bibinfo {pages} {373--377}
  (\bibinfo {year} {2021})}\BibitemShut {NoStop}%
\bibitem [{\citenamefont {Yin}\ \emph {et~al.}(2013)\citenamefont {Yin},
  \citenamefont {Geraci},\ and\ \citenamefont {Li}}]{yin2013optomechanics}%
  \BibitemOpen
  \bibfield  {author} {\bibinfo {author} {\bibfnamefont {Z.-Q.}\ \bibnamefont
  {Yin}}, \bibinfo {author} {\bibfnamefont {A.~A.}\ \bibnamefont {Geraci}}, \
  and\ \bibinfo {author} {\bibfnamefont {T.}~\bibnamefont {Li}},\ }\bibfield
  {title} {\enquote {\bibinfo {title} {Optomechanics of levitated dielectric
  particles},}\ }\href@noop {} {\bibfield  {journal} {\bibinfo  {journal}
  {International Journal of Modern Physics B}\ }\textbf {\bibinfo {volume}
  {27}},\ \bibinfo {pages} {1330018} (\bibinfo {year} {2013})}\BibitemShut
  {NoStop}%
\bibitem [{\citenamefont {Millen}\ \emph {et~al.}(2020)\citenamefont {Millen},
  \citenamefont {Monteiro}, \citenamefont {Pettit},\ and\ \citenamefont
  {Vamivakas}}]{millen2020optomechanics}%
  \BibitemOpen
  \bibfield  {author} {\bibinfo {author} {\bibfnamefont {J.}~\bibnamefont
  {Millen}}, \bibinfo {author} {\bibfnamefont {T.~S.}\ \bibnamefont
  {Monteiro}}, \bibinfo {author} {\bibfnamefont {R.}~\bibnamefont {Pettit}}, \
  and\ \bibinfo {author} {\bibfnamefont {A.~N.}\ \bibnamefont {Vamivakas}},\
  }\bibfield  {title} {\enquote {\bibinfo {title} {Optomechanics with levitated
  particles},}\ }\href@noop {} {\bibfield  {journal} {\bibinfo  {journal}
  {Reports on Progress in Physics}\ }\textbf {\bibinfo {volume} {83}},\
  \bibinfo {pages} {026401} (\bibinfo {year} {2020})}\BibitemShut {NoStop}%
\bibitem [{\citenamefont {Moore}\ and\ \citenamefont
  {Geraci}(2021)}]{moore2021searching}%
  \BibitemOpen
  \bibfield  {author} {\bibinfo {author} {\bibfnamefont {D.~C.}\ \bibnamefont
  {Moore}}\ and\ \bibinfo {author} {\bibfnamefont {A.~A.}\ \bibnamefont
  {Geraci}},\ }\bibfield  {title} {\enquote {\bibinfo {title} {Searching for
  new physics using optically levitated sensors},}\ }\href@noop {} {\bibfield
  {journal} {\bibinfo  {journal} {Quantum Science and Technology}\ }\textbf
  {\bibinfo {volume} {6}},\ \bibinfo {pages} {014008} (\bibinfo {year}
  {2021})}\BibitemShut {NoStop}%
\bibitem [{\citenamefont {Jayich}\ \emph {et~al.}(2008)\citenamefont {Jayich},
  \citenamefont {Sankey}, \citenamefont {Zwickl}, \citenamefont {Yang},
  \citenamefont {Thompson}, \citenamefont {Girvin}, \citenamefont {Clerk},
  \citenamefont {Marquardt},\ and\ \citenamefont
  {Harris}}]{jayich2008dispersive}%
  \BibitemOpen
  \bibfield  {author} {\bibinfo {author} {\bibfnamefont {A.}~\bibnamefont
  {Jayich}}, \bibinfo {author} {\bibfnamefont {J.}~\bibnamefont {Sankey}},
  \bibinfo {author} {\bibfnamefont {B.}~\bibnamefont {Zwickl}}, \bibinfo
  {author} {\bibfnamefont {C.}~\bibnamefont {Yang}}, \bibinfo {author}
  {\bibfnamefont {J.}~\bibnamefont {Thompson}}, \bibinfo {author}
  {\bibfnamefont {S.}~\bibnamefont {Girvin}}, \bibinfo {author} {\bibfnamefont
  {A.}~\bibnamefont {Clerk}}, \bibinfo {author} {\bibfnamefont
  {F.}~\bibnamefont {Marquardt}}, \ and\ \bibinfo {author} {\bibfnamefont
  {J.}~\bibnamefont {Harris}},\ }\bibfield  {title} {\enquote {\bibinfo {title}
  {Dispersive optomechanics: a membrane inside a cavity},}\ }\href@noop {}
  {\bibfield  {journal} {\bibinfo  {journal} {New Journal of Physics}\ }\textbf
  {\bibinfo {volume} {10}},\ \bibinfo {pages} {095008} (\bibinfo {year}
  {2008})}\BibitemShut {NoStop}%
\bibitem [{\citenamefont {Bunch}\ \emph {et~al.}(2007)\citenamefont {Bunch},
  \citenamefont {Van Der~Zande}, \citenamefont {Verbridge}, \citenamefont
  {Frank}, \citenamefont {Tanenbaum}, \citenamefont {Parpia}, \citenamefont
  {Craighead},\ and\ \citenamefont {McEuen}}]{bunch2007electromechanical}%
  \BibitemOpen
  \bibfield  {author} {\bibinfo {author} {\bibfnamefont {J.~S.}\ \bibnamefont
  {Bunch}}, \bibinfo {author} {\bibfnamefont {A.~M.}\ \bibnamefont {Van
  Der~Zande}}, \bibinfo {author} {\bibfnamefont {S.~S.}\ \bibnamefont
  {Verbridge}}, \bibinfo {author} {\bibfnamefont {I.~W.}\ \bibnamefont
  {Frank}}, \bibinfo {author} {\bibfnamefont {D.~M.}\ \bibnamefont
  {Tanenbaum}}, \bibinfo {author} {\bibfnamefont {J.~M.}\ \bibnamefont
  {Parpia}}, \bibinfo {author} {\bibfnamefont {H.~G.}\ \bibnamefont
  {Craighead}}, \ and\ \bibinfo {author} {\bibfnamefont {P.~L.}\ \bibnamefont
  {McEuen}},\ }\bibfield  {title} {\enquote {\bibinfo {title}
  {Electromechanical resonators from graphene sheets},}\ }\href@noop {}
  {\bibfield  {journal} {\bibinfo  {journal} {Science}\ }\textbf {\bibinfo
  {volume} {315}},\ \bibinfo {pages} {490--493} (\bibinfo {year}
  {2007})}\BibitemShut {NoStop}%
\bibitem [{\citenamefont {Zwickl}\ \emph {et~al.}(2008)\citenamefont {Zwickl},
  \citenamefont {Shanks}, \citenamefont {Jayich}, \citenamefont {Yang},
  \citenamefont {Bleszynski~Jayich}, \citenamefont {Thompson},\ and\
  \citenamefont {Harris}}]{zwickl2008high}%
  \BibitemOpen
  \bibfield  {author} {\bibinfo {author} {\bibfnamefont {B.}~\bibnamefont
  {Zwickl}}, \bibinfo {author} {\bibfnamefont {W.}~\bibnamefont {Shanks}},
  \bibinfo {author} {\bibfnamefont {A.}~\bibnamefont {Jayich}}, \bibinfo
  {author} {\bibfnamefont {C.}~\bibnamefont {Yang}}, \bibinfo {author}
  {\bibfnamefont {A.}~\bibnamefont {Bleszynski~Jayich}}, \bibinfo {author}
  {\bibfnamefont {J.}~\bibnamefont {Thompson}}, \ and\ \bibinfo {author}
  {\bibfnamefont {J.}~\bibnamefont {Harris}},\ }\bibfield  {title} {\enquote
  {\bibinfo {title} {High quality mechanical and optical properties of
  commercial silicon nitride membranes},}\ }\href@noop {} {\bibfield  {journal}
  {\bibinfo  {journal} {Applied Physics Letters}\ }\textbf {\bibinfo {volume}
  {92}},\ \bibinfo {pages} {103125} (\bibinfo {year} {2008})}\BibitemShut
  {NoStop}%
\bibitem [{\citenamefont {Wilson}\ \emph {et~al.}(2009)\citenamefont {Wilson},
  \citenamefont {Regal}, \citenamefont {Papp},\ and\ \citenamefont
  {Kimble}}]{wilson2009cavity}%
  \BibitemOpen
  \bibfield  {author} {\bibinfo {author} {\bibfnamefont {D.~J.}\ \bibnamefont
  {Wilson}}, \bibinfo {author} {\bibfnamefont {C.~A.}\ \bibnamefont {Regal}},
  \bibinfo {author} {\bibfnamefont {S.~B.}\ \bibnamefont {Papp}}, \ and\
  \bibinfo {author} {\bibfnamefont {H.}~\bibnamefont {Kimble}},\ }\bibfield
  {title} {\enquote {\bibinfo {title} {Cavity optomechanics with stoichiometric
  sin films},}\ }\href@noop {} {\bibfield  {journal} {\bibinfo  {journal}
  {Physical review letters}\ }\textbf {\bibinfo {volume} {103}},\ \bibinfo
  {pages} {207204} (\bibinfo {year} {2009})}\BibitemShut {NoStop}%
\bibitem [{\citenamefont {Vitali}\ \emph {et~al.}(2001)\citenamefont {Vitali},
  \citenamefont {Mancini},\ and\ \citenamefont
  {Tombesi}}]{vitali2001optomechanical}%
  \BibitemOpen
  \bibfield  {author} {\bibinfo {author} {\bibfnamefont {D.}~\bibnamefont
  {Vitali}}, \bibinfo {author} {\bibfnamefont {S.}~\bibnamefont {Mancini}}, \
  and\ \bibinfo {author} {\bibfnamefont {P.}~\bibnamefont {Tombesi}},\
  }\bibfield  {title} {\enquote {\bibinfo {title} {Optomechanical scheme for
  the detection of weak impulsive forces},}\ }\href@noop {} {\bibfield
  {journal} {\bibinfo  {journal} {Physical Review A}\ }\textbf {\bibinfo
  {volume} {64}},\ \bibinfo {pages} {051401} (\bibinfo {year}
  {2001})}\BibitemShut {NoStop}%
\bibitem [{\citenamefont {Harris}\ \emph {et~al.}(2013)\citenamefont {Harris},
  \citenamefont {McAuslan}, \citenamefont {Stace}, \citenamefont {Doherty},\
  and\ \citenamefont {Bowen}}]{harris2013minimum}%
  \BibitemOpen
  \bibfield  {author} {\bibinfo {author} {\bibfnamefont {G.~I.}\ \bibnamefont
  {Harris}}, \bibinfo {author} {\bibfnamefont {D.~L.}\ \bibnamefont
  {McAuslan}}, \bibinfo {author} {\bibfnamefont {T.~M.}\ \bibnamefont {Stace}},
  \bibinfo {author} {\bibfnamefont {A.~C.}\ \bibnamefont {Doherty}}, \ and\
  \bibinfo {author} {\bibfnamefont {W.~P.}\ \bibnamefont {Bowen}},\ }\bibfield
  {title} {\enquote {\bibinfo {title} {Minimum requirements for feedback
  enhanced force sensing},}\ }\href@noop {} {\bibfield  {journal} {\bibinfo
  {journal} {Physical review letters}\ }\textbf {\bibinfo {volume} {111}},\
  \bibinfo {pages} {103603} (\bibinfo {year} {2013})}\BibitemShut {NoStop}%
\bibitem [{\citenamefont {Yu}\ \emph {et~al.}(2014)\citenamefont {Yu},
  \citenamefont {Cicak}, \citenamefont {Kampel}, \citenamefont {Tsaturyan},
  \citenamefont {Purdy}, \citenamefont {Simmonds},\ and\ \citenamefont
  {Regal}}]{yu2014phononic}%
  \BibitemOpen
  \bibfield  {author} {\bibinfo {author} {\bibfnamefont {P.-L.}\ \bibnamefont
  {Yu}}, \bibinfo {author} {\bibfnamefont {K.}~\bibnamefont {Cicak}}, \bibinfo
  {author} {\bibfnamefont {N.}~\bibnamefont {Kampel}}, \bibinfo {author}
  {\bibfnamefont {Y.}~\bibnamefont {Tsaturyan}}, \bibinfo {author}
  {\bibfnamefont {T.}~\bibnamefont {Purdy}}, \bibinfo {author} {\bibfnamefont
  {R.}~\bibnamefont {Simmonds}}, \ and\ \bibinfo {author} {\bibfnamefont
  {C.}~\bibnamefont {Regal}},\ }\bibfield  {title} {\enquote {\bibinfo {title}
  {A phononic bandgap shield for high-q membrane microresonators},}\
  }\href@noop {} {\bibfield  {journal} {\bibinfo  {journal} {Applied Physics
  Letters}\ }\textbf {\bibinfo {volume} {104}},\ \bibinfo {pages} {023510}
  (\bibinfo {year} {2014})}\BibitemShut {NoStop}%
\bibitem [{\citenamefont {Kirchhof}\ \emph {et~al.}(2021)\citenamefont
  {Kirchhof}, \citenamefont {Weinel}, \citenamefont {Heeg}, \citenamefont
  {Deinhart}, \citenamefont {Kovalchuk}, \citenamefont {H{\"{o}}flich},\ and\
  \citenamefont {Bolotin}}]{Kirchhof2021}%
  \BibitemOpen
  \bibfield  {author} {\bibinfo {author} {\bibfnamefont {J.~N.}\ \bibnamefont
  {Kirchhof}}, \bibinfo {author} {\bibfnamefont {K.}~\bibnamefont {Weinel}},
  \bibinfo {author} {\bibfnamefont {S.}~\bibnamefont {Heeg}}, \bibinfo {author}
  {\bibfnamefont {V.}~\bibnamefont {Deinhart}}, \bibinfo {author}
  {\bibfnamefont {S.}~\bibnamefont {Kovalchuk}}, \bibinfo {author}
  {\bibfnamefont {K.}~\bibnamefont {H{\"{o}}flich}}, \ and\ \bibinfo {author}
  {\bibfnamefont {K.~I.}\ \bibnamefont {Bolotin}},\ }\bibfield  {title}
  {\enquote {\bibinfo {title} {{Tunable Graphene Phononic Crystal}},}\ }\href
  {\doibase 10.1021/acs.nanolett.0c04986} {\bibfield  {journal} {\bibinfo
  {journal} {Nano Lett.}\ }\textbf {\bibinfo {volume} {21}},\ \bibinfo {pages}
  {2174} (\bibinfo {year} {2021})}\BibitemShut {NoStop}%
\bibitem [{\citenamefont {Knapen}\ \emph {et~al.}(2018)\citenamefont {Knapen},
  \citenamefont {Lin}, \citenamefont {Pyle},\ and\ \citenamefont
  {Zurek}}]{Knapen:2017ekk}%
  \BibitemOpen
  \bibfield  {author} {\bibinfo {author} {\bibfnamefont {S.}~\bibnamefont
  {Knapen}}, \bibinfo {author} {\bibfnamefont {T.}~\bibnamefont {Lin}},
  \bibinfo {author} {\bibfnamefont {M.}~\bibnamefont {Pyle}}, \ and\ \bibinfo
  {author} {\bibfnamefont {K.~M.}\ \bibnamefont {Zurek}},\ }\bibfield  {title}
  {\enquote {\bibinfo {title} {{Detection of Light Dark Matter With Optical
  Phonons in Polar Materials}},}\ }\href {\doibase
  10.1016/j.physletb.2018.08.064} {\bibfield  {journal} {\bibinfo  {journal}
  {Phys. Lett. B}\ }\textbf {\bibinfo {volume} {785}},\ \bibinfo {pages}
  {386--390} (\bibinfo {year} {2018})},\ \Eprint
  {http://arxiv.org/abs/1712.06598} {arXiv:1712.06598 [hep-ph]} \BibitemShut
  {NoStop}%
\bibitem [{\citenamefont {Ramsay}(1956)}]{RamsayMolBeams}%
  \BibitemOpen
  \bibfield  {author} {\bibinfo {author} {\bibfnamefont {N.~F.}\ \bibnamefont
  {Ramsay}},\ }\href@noop {} {\emph {\bibinfo {title} {Molecular Beams}}}\
  (\bibinfo  {publisher} {Oxford University Press Inc.},\ \bibinfo {address}
  {New York},\ \bibinfo {year} {1956})\BibitemShut {NoStop}%
\bibitem [{\citenamefont {Cavalleri}\ \emph {et~al.}(2010)\citenamefont
  {Cavalleri}, \citenamefont {Ciani}, \citenamefont {Dolesi}, \citenamefont
  {Hueller}, \citenamefont {Nicolodi}, \citenamefont {Tombolato}, \citenamefont
  {Vitale}, \citenamefont {Wass},\ and\ \citenamefont {Weber}}]{Cavalleri2010}%
  \BibitemOpen
  \bibfield  {author} {\bibinfo {author} {\bibfnamefont {A.}~\bibnamefont
  {Cavalleri}}, \bibinfo {author} {\bibfnamefont {G.}~\bibnamefont {Ciani}},
  \bibinfo {author} {\bibfnamefont {R.}~\bibnamefont {Dolesi}}, \bibinfo
  {author} {\bibfnamefont {M.}~\bibnamefont {Hueller}}, \bibinfo {author}
  {\bibfnamefont {D.}~\bibnamefont {Nicolodi}}, \bibinfo {author}
  {\bibfnamefont {D.}~\bibnamefont {Tombolato}}, \bibinfo {author}
  {\bibfnamefont {S.}~\bibnamefont {Vitale}}, \bibinfo {author} {\bibfnamefont
  {P.~J.}\ \bibnamefont {Wass}}, \ and\ \bibinfo {author} {\bibfnamefont
  {W.~J.}\ \bibnamefont {Weber}},\ }\bibfield  {title} {\enquote {\bibinfo
  {title} {{Gas damping force noise on a macroscopic test body in an infinite
  gas reservoir}},}\ }\href {\doibase 10.1016/j.physleta.2010.06.041}
  {\bibfield  {journal} {\bibinfo  {journal} {Phys. Lett. A}\ }\textbf
  {\bibinfo {volume} {374}},\ \bibinfo {pages} {3365} (\bibinfo {year}
  {2010})}\BibitemShut {NoStop}%
\bibitem [{\citenamefont {Martinetz}\ \emph {et~al.}(2018)\citenamefont
  {Martinetz}, \citenamefont {Hornberger},\ and\ \citenamefont
  {Stickler}}]{Martinetz2018}%
  \BibitemOpen
  \bibfield  {author} {\bibinfo {author} {\bibfnamefont {L.}~\bibnamefont
  {Martinetz}}, \bibinfo {author} {\bibfnamefont {K.}~\bibnamefont
  {Hornberger}}, \ and\ \bibinfo {author} {\bibfnamefont {B.~A.}\ \bibnamefont
  {Stickler}},\ }\bibfield  {title} {\enquote {\bibinfo {title} {{Gas-induced
  friction and diffusion of rigid rotors}},}\ }\href {\doibase
  10.1103/PhysRevE.97.052112} {\bibfield  {journal} {\bibinfo  {journal} {Phys.
  Rev. E}\ }\textbf {\bibinfo {volume} {97}},\ \bibinfo {pages} {052112}
  (\bibinfo {year} {2018})}\BibitemShut {NoStop}%
\bibitem [{\citenamefont {Blakemore}\ \emph {et~al.}(2020)\citenamefont
  {Blakemore}, \citenamefont {Martin}, \citenamefont {Fieguth}, \citenamefont
  {Kawasaki}, \citenamefont {Priel}, \citenamefont {Rider},\ and\ \citenamefont
  {Gratta}}]{Blakemore2020}%
  \BibitemOpen
  \bibfield  {author} {\bibinfo {author} {\bibfnamefont {C.~P.}\ \bibnamefont
  {Blakemore}}, \bibinfo {author} {\bibfnamefont {D.}~\bibnamefont {Martin}},
  \bibinfo {author} {\bibfnamefont {A.}~\bibnamefont {Fieguth}}, \bibinfo
  {author} {\bibfnamefont {A.}~\bibnamefont {Kawasaki}}, \bibinfo {author}
  {\bibfnamefont {N.}~\bibnamefont {Priel}}, \bibinfo {author} {\bibfnamefont
  {A.~D.}\ \bibnamefont {Rider}}, \ and\ \bibinfo {author} {\bibfnamefont
  {G.}~\bibnamefont {Gratta}},\ }\bibfield  {title} {\enquote {\bibinfo {title}
  {{Absolute pressure and gas species identification with an optically
  levitated rotor}},}\ }\href {\doibase 10.1116/1.5139638} {\bibfield
  {journal} {\bibinfo  {journal} {J. Vac. Sci. Technol. B}\ }\textbf {\bibinfo
  {volume} {38}},\ \bibinfo {pages} {024201} (\bibinfo {year}
  {2020})}\BibitemShut {NoStop}%
\bibitem [{\citenamefont {Fremerey}(1985)}]{Fremerey1985}%
  \BibitemOpen
  \bibfield  {author} {\bibinfo {author} {\bibfnamefont {J.~K.}\ \bibnamefont
  {Fremerey}},\ }\bibfield  {title} {\enquote {\bibinfo {title} {{The spinning
  rotor gauge}},}\ }\href@noop {} {\bibfield  {journal} {\bibinfo  {journal}
  {J. Vac. Sci. Technol. A}\ }\textbf {\bibinfo {volume} {3}},\ \bibinfo
  {pages} {1715} (\bibinfo {year} {1985})}\BibitemShut {NoStop}%
\bibitem [{\citenamefont {Tilford}(1992)}]{Tilford1992}%
  \BibitemOpen
  \bibfield  {author} {\bibinfo {author} {\bibfnamefont {C.~R.}\ \bibnamefont
  {Tilford}},\ }\bibfield  {title} {\enquote {\bibinfo {title} {{Pressure and
  Vacuum Measurements}},}\ }in\ \href@noop {} {\emph {\bibinfo {booktitle}
  {Phys. Methods Chem.}}},\ Vol.~\bibinfo {volume} {6},\ \bibinfo {editor}
  {edited by\ \bibinfo {editor} {\bibfnamefont {B.~W.}\ \bibnamefont
  {Rossiter}}\ and\ \bibinfo {editor} {\bibfnamefont {R.~C.}\ \bibnamefont
  {Baetzold}}}\ (\bibinfo  {publisher} {John Wiley and Sons, Inc.},\ \bibinfo
  {year} {1992})\ \bibinfo {edition} {2nd}\ ed.,\ pp.\ \bibinfo {pages}
  {101--173}\BibitemShut {NoStop}%
\bibitem [{\citenamefont {Scherschligt}\ \emph {et~al.}(2018)\citenamefont
  {Scherschligt}, \citenamefont {Fedchak}, \citenamefont {Ahmed}, \citenamefont
  {Barker}, \citenamefont {Douglass}, \citenamefont {Eckel}, \citenamefont
  {Hanson}, \citenamefont {Hendricks}, \citenamefont {Klimov}, \citenamefont
  {Purdy}, \citenamefont {Ricker}, \citenamefont {Singh},\ and\ \citenamefont
  {Stone}}]{Scherschligt2018}%
  \BibitemOpen
  \bibfield  {author} {\bibinfo {author} {\bibfnamefont {J.}~\bibnamefont
  {Scherschligt}}, \bibinfo {author} {\bibfnamefont {J.~A.}\ \bibnamefont
  {Fedchak}}, \bibinfo {author} {\bibfnamefont {Z.}~\bibnamefont {Ahmed}},
  \bibinfo {author} {\bibfnamefont {D.~S.}\ \bibnamefont {Barker}}, \bibinfo
  {author} {\bibfnamefont {K.}~\bibnamefont {Douglass}}, \bibinfo {author}
  {\bibfnamefont {S.}~\bibnamefont {Eckel}}, \bibinfo {author} {\bibfnamefont
  {E.}~\bibnamefont {Hanson}}, \bibinfo {author} {\bibfnamefont
  {J.}~\bibnamefont {Hendricks}}, \bibinfo {author} {\bibfnamefont
  {N.}~\bibnamefont {Klimov}}, \bibinfo {author} {\bibfnamefont
  {T.}~\bibnamefont {Purdy}}, \bibinfo {author} {\bibfnamefont
  {J.}~\bibnamefont {Ricker}}, \bibinfo {author} {\bibfnamefont
  {R.}~\bibnamefont {Singh}}, \ and\ \bibinfo {author} {\bibfnamefont
  {J.}~\bibnamefont {Stone}},\ }\bibfield  {title} {\enquote {\bibinfo {title}
  {{Review Article: Quantum-based vacuum metrology at the National Institute of
  Standards and Technology}},}\ }\href {\doibase 10.1116/1.5033568} {\bibfield
  {journal} {\bibinfo  {journal} {J. Vac. Sci. Technol. A}\ }\textbf {\bibinfo
  {volume} {36}},\ \bibinfo {pages} {040801} (\bibinfo {year}
  {2018})}\BibitemShut {NoStop}%
\bibitem [{\citenamefont {Hsu}\ \emph {et~al.}(2018)\citenamefont {Hsu},
  \citenamefont {Ramakrishna}, \citenamefont {Zanini}, \citenamefont
  {Spencer},\ and\ \citenamefont {Isa}}]{Hsu2018}%
  \BibitemOpen
  \bibfield  {author} {\bibinfo {author} {\bibfnamefont {C.~P.}\ \bibnamefont
  {Hsu}}, \bibinfo {author} {\bibfnamefont {S.~N.}\ \bibnamefont
  {Ramakrishna}}, \bibinfo {author} {\bibfnamefont {M.}~\bibnamefont {Zanini}},
  \bibinfo {author} {\bibfnamefont {N.~D.}\ \bibnamefont {Spencer}}, \ and\
  \bibinfo {author} {\bibfnamefont {L.}~\bibnamefont {Isa}},\ }\bibfield
  {title} {\enquote {\bibinfo {title} {{Roughness-dependent tribology effects
  on discontinuous shear thickening}},}\ }\href {\doibase
  10.1073/pnas.1801066115} {\bibfield  {journal} {\bibinfo  {journal} {Proc.
  Natl. Acad. Sci. U. S. A.}\ }\textbf {\bibinfo {volume} {115}},\ \bibinfo
  {pages} {5117} (\bibinfo {year} {2018})}\BibitemShut {NoStop}%
\bibitem [{\citenamefont {Lui}\ \emph {et~al.}(2009)\citenamefont {Lui},
  \citenamefont {Liu}, \citenamefont {Mak}, \citenamefont {Flynn},\ and\
  \citenamefont {Heinz}}]{Lui2009}%
  \BibitemOpen
  \bibfield  {author} {\bibinfo {author} {\bibfnamefont {C.~H.}\ \bibnamefont
  {Lui}}, \bibinfo {author} {\bibfnamefont {L.}~\bibnamefont {Liu}}, \bibinfo
  {author} {\bibfnamefont {K.~F.}\ \bibnamefont {Mak}}, \bibinfo {author}
  {\bibfnamefont {G.~W.}\ \bibnamefont {Flynn}}, \ and\ \bibinfo {author}
  {\bibfnamefont {T.~F.}\ \bibnamefont {Heinz}},\ }\bibfield  {title} {\enquote
  {\bibinfo {title} {{Ultraflat graphene}},}\ }\href {\doibase
  10.1038/nature08569} {\bibfield  {journal} {\bibinfo  {journal} {Nature}\
  }\textbf {\bibinfo {volume} {462}},\ \bibinfo {pages} {339} (\bibinfo {year}
  {2009})}\BibitemShut {NoStop}%
\bibitem [{\citenamefont {Ehinger}\ \emph {et~al.}(2022)\citenamefont
  {Ehinger}, \citenamefont {Acharya}, \citenamefont {Barker}, \citenamefont
  {Fedchak}, \citenamefont {Scherschligt}, \citenamefont {Tiesinga},\ and\
  \citenamefont {Eckel}}]{Ehinger2022}%
  \BibitemOpen
  \bibfield  {author} {\bibinfo {author} {\bibfnamefont {L.~H.}\ \bibnamefont
  {Ehinger}}, \bibinfo {author} {\bibfnamefont {B.~P.}\ \bibnamefont
  {Acharya}}, \bibinfo {author} {\bibfnamefont {D.~S.}\ \bibnamefont {Barker}},
  \bibinfo {author} {\bibfnamefont {J.~A.}\ \bibnamefont {Fedchak}}, \bibinfo
  {author} {\bibfnamefont {J.}~\bibnamefont {Scherschligt}}, \bibinfo {author}
  {\bibfnamefont {E.}~\bibnamefont {Tiesinga}}, \ and\ \bibinfo {author}
  {\bibfnamefont {S.}~\bibnamefont {Eckel}},\ }\bibfield  {title} {\enquote
  {\bibinfo {title} {{Comparison of two multiplexed portable cold-atom vacuum
  standards}},}\ }\href {\doibase 10.1116/5.0095011} {\bibfield  {journal}
  {\bibinfo  {journal} {AVS Quantum Sci.}\ }\textbf {\bibinfo {volume} {4}},\
  \bibinfo {pages} {034403} (\bibinfo {year} {2022})}\BibitemShut {NoStop}%
\bibitem [{\citenamefont {Blakemore}\ \emph {et~al.}(2019)\citenamefont
  {Blakemore}, \citenamefont {Rider}, \citenamefont {Roy}, \citenamefont
  {Fieguth}, \citenamefont {Kawasaki}, \citenamefont {Priel},\ and\
  \citenamefont {Gratta}}]{Blakemore2019}%
  \BibitemOpen
  \bibfield  {author} {\bibinfo {author} {\bibfnamefont {C.~P.}\ \bibnamefont
  {Blakemore}}, \bibinfo {author} {\bibfnamefont {A.~D.}\ \bibnamefont
  {Rider}}, \bibinfo {author} {\bibfnamefont {S.}~\bibnamefont {Roy}}, \bibinfo
  {author} {\bibfnamefont {A.}~\bibnamefont {Fieguth}}, \bibinfo {author}
  {\bibfnamefont {A.}~\bibnamefont {Kawasaki}}, \bibinfo {author}
  {\bibfnamefont {N.}~\bibnamefont {Priel}}, \ and\ \bibinfo {author}
  {\bibfnamefont {G.}~\bibnamefont {Gratta}},\ }\bibfield  {title} {\enquote
  {\bibinfo {title} {{Precision mass and density measurement of individual
  optically levitated microspheres}},}\ }\href {\doibase
  10.1103/PhysRevApplied.12.024037} {\bibfield  {journal} {\bibinfo  {journal}
  {Phys. Rev. Appl.}\ }\textbf {\bibinfo {volume} {12}},\ \bibinfo {pages}
  {024037} (\bibinfo {year} {2019})}\BibitemShut {NoStop}%
\bibitem [{\citenamefont {Purdy}\ \emph {et~al.}(2017)\citenamefont {Purdy},
  \citenamefont {Grutter}, \citenamefont {Srinivasan},\ and\ \citenamefont
  {Taylor}}]{Purdy2017}%
  \BibitemOpen
  \bibfield  {author} {\bibinfo {author} {\bibfnamefont {T.~P.}\ \bibnamefont
  {Purdy}}, \bibinfo {author} {\bibfnamefont {K.~E.}\ \bibnamefont {Grutter}},
  \bibinfo {author} {\bibfnamefont {K.}~\bibnamefont {Srinivasan}}, \ and\
  \bibinfo {author} {\bibfnamefont {J.~M.}\ \bibnamefont {Taylor}},\ }\bibfield
   {title} {\enquote {\bibinfo {title} {{Quantum correlations from a
  room-temperature optomechanical cavity}},}\ }\href {\doibase
  10.1126/science.aag1407} {\bibfield  {journal} {\bibinfo  {journal}
  {Science}\ }\textbf {\bibinfo {volume} {356}},\ \bibinfo {pages} {1265}
  (\bibinfo {year} {2017})}\BibitemShut {NoStop}%
\bibitem [{\citenamefont {Hebestreit}\ \emph {et~al.}(2018)\citenamefont
  {Hebestreit}, \citenamefont {Reimann}, \citenamefont {Frimmer},\ and\
  \citenamefont {Novotny}}]{Hebestreit2018}%
  \BibitemOpen
  \bibfield  {author} {\bibinfo {author} {\bibfnamefont {E.}~\bibnamefont
  {Hebestreit}}, \bibinfo {author} {\bibfnamefont {R.}~\bibnamefont {Reimann}},
  \bibinfo {author} {\bibfnamefont {M.}~\bibnamefont {Frimmer}}, \ and\
  \bibinfo {author} {\bibfnamefont {L.}~\bibnamefont {Novotny}},\ }\bibfield
  {title} {\enquote {\bibinfo {title} {{Measuring the internal temperature of a
  levitated nanoparticle in high vacuum}},}\ }\href {\doibase
  10.1103/PhysRevA.97.043803} {\bibfield  {journal} {\bibinfo  {journal} {Phys.
  Rev. A}\ }\textbf {\bibinfo {volume} {97}},\ \bibinfo {pages} {043803}
  (\bibinfo {year} {2018})}\BibitemShut {NoStop}%
\bibitem [{\citenamefont {Vinante}\ \emph {et~al.}(2019)\citenamefont
  {Vinante}, \citenamefont {Pontin}, \citenamefont {Rashid}, \citenamefont
  {Toro{\v{s}}}, \citenamefont {Barker},\ and\ \citenamefont
  {Ulbricht}}]{Vinante2019}%
  \BibitemOpen
  \bibfield  {author} {\bibinfo {author} {\bibfnamefont {A.}~\bibnamefont
  {Vinante}}, \bibinfo {author} {\bibfnamefont {A.}~\bibnamefont {Pontin}},
  \bibinfo {author} {\bibfnamefont {M.}~\bibnamefont {Rashid}}, \bibinfo
  {author} {\bibfnamefont {M.}~\bibnamefont {Toro{\v{s}}}}, \bibinfo {author}
  {\bibfnamefont {P.~F.}\ \bibnamefont {Barker}}, \ and\ \bibinfo {author}
  {\bibfnamefont {H.}~\bibnamefont {Ulbricht}},\ }\bibfield  {title} {\enquote
  {\bibinfo {title} {{Testing collapse models with levitated nanoparticles:
  Detection challenge}},}\ }\href {\doibase 10.1103/PhysRevA.100.012119}
  {\bibfield  {journal} {\bibinfo  {journal} {Phys. Rev. A}\ }\textbf {\bibinfo
  {volume} {100}},\ \bibinfo {pages} {012119} (\bibinfo {year}
  {2019})}\BibitemShut {NoStop}%
\bibitem [{\citenamefont {Monteiro}\ \emph {et~al.}(2020)\citenamefont
  {Monteiro}, \citenamefont {Afek}, \citenamefont {Carney}, \citenamefont
  {Krnjaic}, \citenamefont {Wang},\ and\ \citenamefont
  {Moore}}]{monteiro2020search}%
  \BibitemOpen
  \bibfield  {author} {\bibinfo {author} {\bibfnamefont {F.}~\bibnamefont
  {Monteiro}}, \bibinfo {author} {\bibfnamefont {G.}~\bibnamefont {Afek}},
  \bibinfo {author} {\bibfnamefont {D.}~\bibnamefont {Carney}}, \bibinfo
  {author} {\bibfnamefont {G.}~\bibnamefont {Krnjaic}}, \bibinfo {author}
  {\bibfnamefont {J.}~\bibnamefont {Wang}}, \ and\ \bibinfo {author}
  {\bibfnamefont {D.~C.}\ \bibnamefont {Moore}},\ }\bibfield  {title} {\enquote
  {\bibinfo {title} {Search for composite dark matter with optically levitated
  sensors},}\ }\href@noop {} {\bibfield  {journal} {\bibinfo  {journal}
  {Physical Review Letters}\ }\textbf {\bibinfo {volume} {125}},\ \bibinfo
  {pages} {181102} (\bibinfo {year} {2020})}\BibitemShut {NoStop}%
\end{thebibliography}%

\clearpage

\appendix

\onecolumngrid
\section{Collision spectrum calculation}
\label{appendix-gas}

We determine the collision spectrum using the kinetic theory of gases.
The number of molecular collisions with the sensor surface element $dA$ in time element $dt$ that have incoming velocity $\vec{v}_i$ and outgoing velocity $\vec{v}_o$ is~\cite{Cavalleri2010, Martinetz2018}
\begin{equation}
    \begin{split}
    \label{eq:colls}
        d^8N_c(\vec{v}_i,\vec{v}_o) & = n_g\,dA\,dt \bigg(\frac{1}{2\pi \overline{v}^2}\bigg)^{3/2} v_i \,\mathrm{cos} \,\theta_i \,e^{-v_i^2\big/2\overline{v}^2} \\
        & \times \frac{1}{2\pi \overline{v}^4} v_o \,\mathrm{cos}\,\theta_o \,\,e^{-v_o^2\big/2\overline{v}^2} \,d\vec{v}_i d\vec{v}_o,
    \end{split}
\end{equation}
where $n_g$ is the gas density, $v_{i,o}$ is the magnitude of $\vec{v}_{i,o}$, $\overline{v} = \sqrt{k_B T/m_g}$ is the root-mean-square thermal velocity of the gas, and $\theta_{i,o}$ is the polar angle between $\vec{v}_{i,o}$ and the surface normal $\hat{u}_\perp$.
Equation~\ref{eq:colls} assumes that the gas molecules scatter diffusely from the sensor surface according to the cosine law after thermalizing with it (see Refs.~\cite{RamsayMolBeams, Cavalleri2010, Martinetz2018, Blakemore2020}) and that the sensor is in thermal equilibrium with the gas. If the sensor is not in thermal equilibrium with the gas, which may occur at low background pressure or high optical power~\cite{Vinante2019}, then the sensor temperature multiplied by the thermal accommodation coefficient should replace the gas temperature on the second line of Eq.~\ref{eq:colls}.

To find the number of collisions that impart momentum $\Delta p$ perpendicular to the surface, we integrate Eq.~\ref{eq:colls} subject to the constraint $\vec{v}_{o}.\hat{u}_\perp+\vec{v}_{i}.\hat{u}_\perp-\Delta p/m_g = 0$.
After transforming to Cartesian coordinates, we have
    \begin{equation}
        \begin{split}
        \label{eq:collscart}
            d^3N_c(\Delta p) & = n_g\,dA\,dt \bigg(\frac{1}{2\pi \overline{v}^2}\bigg)^{3/2} \int_0^{\Delta p/m_g} dv_{i,z} \iint_{-\infty}^\infty dv_{i,x}dv_{i,y} \,v_{i,z} \,e^{-(v_{i,x}^2+v_{i,y}^2+v_{i,z}^2)\big/2\overline{v}^2} \\
            & \times \frac{1}{2\pi \overline{v}^4} dv_{o,z} \iint_{-\infty}^\infty \,dv_{o,x}dv_{o,y} \,v_{o, z} \,e^{-(v_{o,x}^2+v_{o,y}^2+v_{o,z}^2)\big/2\overline{v}^2}, \\
        \end{split}
    \end{equation}
where $v_{i,z}=\vec{v}_i.\hat{z}$ (with $\hat{z}=\hat{u}_\perp$ the unit vector defining the $z$ axis) and so on for the other Cartesian components of $\vec{v}_{i}$ and $\vec{v}_{o}$.
We impose the momentum transfer constraint by taking $v_{o,z}=\Delta p/m_g-v_{i,z}$ and $dv_{o,z}=d\Delta p/m_g$.
Evaluating the integrals over the plane parallel to the surface then yields
    \begin{equation}
        \begin{split}
        \label{eq:colls1d}
            d^3N_c(\Delta p) & = \,\frac{n_g\,dA\,dt}{\overline{v}^2} \bigg(\frac{1}{2\pi \overline{v}^2}\bigg)^{1/2} \frac{d\Delta p}{m_g} \int_0^{\Delta p/m_g} dv_{i,z} \,v_{i,z} (\Delta p/m_g - v_{i, z}) \,e^{-\big(v_{i,z}^2+(\Delta p/m_g-v_{i,z})^2\big)\big/2\overline{v}^2}. \\
        \end{split}
    \end{equation}
The integral in Eq.~\ref{eq:colls1d} can be solved by completing the square and mapping onto known Gaussian integrals.
The result is
    \begin{equation}
        \begin{split}
        \label{eq:integ}
            & \int_0^{\Delta p/m_g} dv_{i,z} \,v_{i,z} (\Delta p/m_g - v_{i, z}) \,e^{-\big(v_{i,z}^2+(\Delta p/m_g-v_{i,z})^2\big)\big/2\overline{v}^2} \\
            = & \,\sqrt{\pi} e^{-\Delta p^2\big/4 m_g^2 \overline{v}^2} \bigg[\frac{1}{\sqrt{\pi}}\Big(\frac{v_{i,z}\overline{v}^2}{2}-\frac{\Delta p \overline{v}^2}{4 m_g}\Big)e^{-\big(v_{i,z}/\overline{v}-\Delta p/2m_g \overline{v}\big)^2} \\
            & + \frac{1}{2}\Big(\frac{\Delta p^2 \overline{v}}{4 m_g^2}-\frac{\overline{v}^3}{2}\Big)\Big(1+\mathrm{erf}\big(v_{i,z}/\overline{v}-\Delta p/2m_g \overline{v}\big)\Big) \bigg]_0^{\Delta p/m_g},
        \end{split}
    \end{equation}
where $\mathrm{erf}$ is the Gaussian error function.
Inserting Eq.~\ref{eq:integ} into Eq.~\ref{eq:colls1d} gives the number of collisions imparting momentum $\Delta p$ per unit area per unit time
    \begin{equation}
        \begin{split}
        \label{eq:collsfin}
            \frac{d^3N_c(\Delta p)}{dA\,dt} = n_g \frac{d\Delta p}{m_g} \bigg(\frac{1}{2\pi \overline{v}^2}\bigg)^{1/2} \bigg( & \frac{\Delta p}{2 m_g}e^{-\Delta p^2\big/2m_g^2\overline{v}^2} \\
            & +\frac{\sqrt{\pi}}{2}\bigg(\frac{\Delta p^2}{2m_g^2\overline{v}}-\overline{v}\bigg)\mathrm{erf}\big(\Delta p/2m_g \overline{v}\big)e^{-\Delta p^2\big/4m_g^2\overline{v}^2}\bigg). \\
        \end{split}
    \end{equation}
If we rearrange Eq.~\ref{eq:collsfin} and integrate over the sensor area, we find the differential event rate
    \begin{equation}
        \begin{split}
        \label{eq:colrate}
            \frac{d\Gamma}{d\Delta p} = & \,\frac{n_g A \Delta p}{4 m_g^2} \bigg(\frac{1}{2\pi \overline{v}^2}\bigg)^{1/2} e^{-\Delta p^2\big/8m_g^2\overline{v}^2}
            \bigg(2 e^{-3\Delta p^2\big/8m_g^2\overline{v}^2} +\frac{\sqrt{\pi}}{2}\bigg(\frac{2 \Delta p}{m_g\overline{v}}-\frac{4m_g\overline{v}}{\Delta p}\bigg)\mathrm{erf}\big(\Delta p/2m_g \overline{v}\big)e^{-\Delta p^2\big/8m_g^2\overline{v}^2}\bigg) \\
            = & \,\frac{n_g A \Delta p}{4 m_g^2} \bigg(\frac{1}{2\pi \overline{v}^2}\bigg)^{1/2} e^{-\Delta p^2\big/8m_g^2\overline{v}^2}\xi\Big(\frac{\Delta p}{m_g\overline{v}}\Big)
        \end{split}
    \end{equation}
where $\Gamma = dN_c/dt$ is the total collision rate and the diffuse scattering correction $\xi(\Delta p/m_g\overline{v})$ is the term in parentheses on the first line of the equation.
Taking $\xi(\Delta p/m_g\overline{v})\rightarrow1$ yields the differential event rate for elastic scattering.
If we note that the Maxwell-Boltzmann distribution for $\Delta p/2 m_g$ is $f_B(\Delta p/2 m_g) = e^{-\Delta p^2\big/8m_g^2\overline{v}^2}/\sqrt{2\pi\overline{v}^2}$ and include momentum accomodation, then Eq.~\ref{eq:colrate} becomes Eq.~\ref{eq:diffeventrate} in the main text.

We calculate the total detectable collision rate by integrating Eq.~\ref{eq:colrate} over $\Delta p$ from $\Delta p_{\rm min}$ to $\infty$, where $\Delta p_{\rm min}$ is the momentum transfer that corresponds to a measurement signal-to-noise ratio of 1 (or a chosen cutoff to ensure no spurious events).
Carrying out the integration yields
    \begin{equation}
        \begin{split}
        \label{eq:qminrate}
            \Gamma\big|_{\Delta p > \Delta p_{\rm min}} = & \,\frac{n_g A}{2} \bigg(\frac{1}{2\pi \overline{v}^2}\bigg)^{1/2}\bigg(\overline{v}^2 e^{-\Delta p_{\rm min}^2/2 m_g^2 \overline{v}^2} \\
            & \qquad +\frac{\sqrt{\pi}}{2}\int_{\Delta p_{\rm min}}^{\infty}\frac{d\Delta p}{m_g}\bigg(\frac{\Delta p^2}{m_g^2\overline{v}}-2\overline{v}\bigg)\mathrm{erf}\big(\Delta p/2m_g \overline{v}\big)e^{-\Delta p^2\big/4m_g^2\overline{v}^2}\bigg). \\
            = & \,\frac{n_g A}{2} \bigg(\frac{1}{2\pi \overline{v}^2}\bigg)^{1/2}\bigg(2 \overline{v}^2 e^{-\Delta p_{\rm min}^2/2 m_g^2 \overline{v}^2} \\
            & \qquad +\frac{\sqrt{\pi}\Delta p_{\rm min}\overline{v}}{m_g}\mathrm{erf}\big(\Delta p_{\rm min}/2m_g \overline{v}\big)e^{-\Delta p_{\rm min}^2\big/4m_g^2\overline{v}^2}\bigg). \\
        \end{split}
    \end{equation}
If we rewrite Eq.~\ref{eq:qminrate} in terms of the expected total collision rate, we get
    \begin{equation}
        \begin{split}
        \label{eq:qminrate2}
            \Gamma\big|_{\Delta p > \Delta p_{\rm min}} = & \,\frac{n_g A \overline{v}}{\sqrt{2\pi}} \bigg(e^{-\Delta p_{\rm min}^2/2 m_g^2 \overline{v}^2} +\frac{\sqrt{\pi}\Delta p_{\rm min}}{2 m_g \overline{v}}\mathrm{erf}\big(\Delta p_{\rm min}/2m_g \overline{v}\big)e^{-\Delta p_{\rm min}^2\big/4m_g^2\overline{v}^2}\bigg). \\
            = & \,\frac{n_g A \overline{v}}{\sqrt{2\pi}} \eta_d(\Delta p_{\rm min}),
        \end{split}
    \end{equation}
where the term in parentheses on the first line defines the detectable momentum cutoff for diffuse scattering $\eta_d(\Delta p_{\rm min}) < 1$.
The detectable momentum cutoff for specular scattering is $\eta_s(\Delta p_{\rm min}) = e^{-\Delta p_{\rm min}^2/8 m_g^2 \overline{v}^2}$, which can be found by taking the $\xi(\Delta p/m_g\overline{v})\rightarrow1$ limit in Eq.~\ref{eq:colrate} and then carrying out the integral in Eq.~\ref{eq:qminrate}.

Equation~\ref{eq:qminrate2} and Eq.~\ref{eq:colrate} assume that all motion perpendicular to the sensor surface is detectable by the readout system.
That is to say, the readout system detects all motion along $\hat{u}_\perp$, which can be the case for tethered devices.
However, motion readout for a levitated sensor will occur along the principle axes of the levitating trap and the details of the experimental setup may prevent simultaneous readout along all three principle axes~\cite{monteiro2020search, magrini2021real}.
To calculate the event rate along a trap axis for a levitated nanosphere, we must project the center-of-mass momentum transfer onto the principle axis of the trap before integrating Eq.~\ref{eq:collsfin} over the sensor surface area.
Taking the readout axis to be $z'$ such that $\hat{u}_\perp.\hat{z}'= \text{cos}\,\theta$ and substituting for $\Delta p$ yields
    \begin{equation}
        \begin{split}
        \label{eq:1dcolrate}
            \frac{d\Gamma}{d\Delta p_{z'}} = & \,\frac{n_g \Delta p_{z'}}{2 m_g^2} \bigg(\frac{1}{2\pi \overline{v}^2}\bigg)^{1/2} 4\pi R^2 \int_0^{\pi/2}\text{sec}^2\theta\, \text{sin}\,\theta\, e^{-\Delta p_{z'}^2\text{sec}^2\theta\big/2m_g^2\overline{v}^2} \\
            & \times \bigg(1 +\frac{\sqrt{\pi}}{2}\bigg(\frac{\Delta p_{z'}\text{sec}\,\theta}{m_g\overline{v}}-\frac{2m_g\overline{v}}{\Delta p_{z'}\text{sec}\,\theta}\bigg)\mathrm{erf}\big(\Delta p_{z'}\text{sec}\,\theta/2m_g \overline{v}\big)e^{\Delta p_{z'}^2\text{sec}^2\theta\big/4m_g^2\overline{v}^2}\bigg) d\theta,
        \end{split}
    \end{equation}
where $R$ is the nanosphere radius and $\Delta p_{z'}$ is the momentum transfer along the $z'$ axis.
To our knowledge, the integral in Eq.~\ref{eq:1dcolrate} does not have an analytic expression. Specifically, the term proportional to \(\text{sec}\,\theta\) must be integrated numerically. However, we can still calculate the total collision rate by integrating over $\Delta p_{z'}$ and switching the order of integration.
The total collision rate is then
    \begin{equation}
        \begin{split}
        \label{eq:1dqminrate}
            \Gamma\big|_{\Delta p_{z'} > \Delta p_{\rm min}} = & \,n_g \sqrt{8\pi} R^2 \overline{v} \int_0^{\pi/2}\bigg(e^{-\Delta p_{\rm min}^2\text{sec}^2\theta/2 m_g^2 \overline{v}^2} \\
            & +\frac{\sqrt{\pi}\Delta p_{\rm min}\text{sec}\,\theta}{2 m_g \overline{v}}\mathrm{erf}\big(\frac{\Delta p_{\rm min}\text{sec}\,\theta}{2m_g \overline{v}}\big)e^{-\Delta p_{\rm min}^2\text{sec}^2\theta\big/4m_g^2\overline{v}^2}\bigg)\text{sin}\,\theta\,d\theta. \\
            = & \,\frac{n_g A \overline{v}}{\sqrt{2\pi}} \eta'_d(\Delta p_{\rm min}),
        \end{split}
    \end{equation}
where the integral term defines the projected momentum cutoff for diffuse scattering $\eta'_d(\Delta p_{\rm min})$.

For specular scattering, the integral for event rate along a single trap axis is analytic, so the event rate is given by
    \begin{equation}
        \begin{split}
        \label{eq:elas1dcolrate}
            \frac{d\Gamma}{d\Delta p_{z'}} = & \,\frac{n_g \Delta p_{z'}}{4 m_g^2} \bigg(\frac{1}{2\pi \overline{v}^2}\bigg)^{1/2} 4\pi R^2 \int_0^{\pi/2}\text{sec}^2\theta\, \text{sin}\,\theta\, e^{-\Delta p_{z'}^2\text{sec}^2\theta\big/8m_g^2\overline{v}^2}d\theta \\
            = & \,-\frac{n_g \Delta p_{z'}}{4 m_g^2} \bigg(\frac{1}{2\pi \overline{v}^2}\bigg)^{1/2} 4\pi R^2 \int_1^0 u^{-2}e^{-\Delta p_{z'}^2\big/8m_g^2\overline{v}^2u^2}du \\
            = & \,\frac{n_g \Delta p_{z'}}{4 m_g^2} \bigg(\frac{1}{2\pi \overline{v}^2}\bigg)^{1/2} 4\pi R^2 \frac{\sqrt{2 \pi} m_g \overline{v}}{\Delta p_{z'}}\mathrm{erfc}\big(\frac{\Delta p_{z'}}{\sqrt{8} m_g \overline{v}}\big) \\
            = & \,\frac{n_g \pi R^2}{m_g}\mathrm{erfc}\big(\frac{\Delta p_{z'}}{\sqrt{8} m_g \overline{v}}\big),
        \end{split}
    \end{equation}
where \(\mathrm{erfc}\) is the complementary Gaussian error function. The total collision rate for specular scattering is then
    \begin{equation}
        \begin{split}
        \label{eq:elas1dqminrate}
            \Gamma\big|_{\Delta p_{z'} > \Delta p_{\rm min}} = & \,\frac{n_g \pi R^2}{m_g}\int_{\Delta p_{\rm min}}^{\infty}\mathrm{erfc}\big(\frac{\Delta p_{z'}}{\sqrt{8} m_g \overline{v}}\big)d\Delta p_{z'} \\
            = & \,\frac{n_g \pi R^2}{m_g}\frac{\sqrt{8}m_g\overline{v}}{\sqrt{\pi}}e^{-\Delta p_{\rm min}^2/8 m_g^2 \overline{v}^2} \\
            = & \,\frac{n_g A \overline{v}}{\sqrt{2\pi}}e^{-\Delta p_{\rm min}^2/8 m_g^2 \overline{v}^2} \\
            = & \,\frac{n_g A \overline{v}}{\sqrt{2\pi}} \eta'_s(\Delta p_{\rm min}),
        \end{split}
    \end{equation}
and we note that \(\eta'_s(\Delta p_{\rm min})=\eta_s(\Delta p_{\rm min})\).

\end{document}